\documentclass[prr,aps,twocolumn,%showpacs,
preprintnumbers,superscriptaddress,longbibliography]{revtex4-2}
\usepackage{amsmath,amssymb,amsfonts,bm}
\usepackage{mathrsfs}
\usepackage{braket}
\usepackage{graphicx}
\usepackage{subcaption}
\usepackage{mathtools}
\usepackage{color}
\usepackage{xcolor}
\usepackage{caption}
\usepackage[colorlinks,linkcolor=violet,citecolor=violet]{hyperref}
\usepackage{comment}
\usepackage{physics}

\newcommand{\T}{\mbox{\tiny T}}
\newcommand{\B}{\mathrm{B}}
\newcommand{\s}{\mbox{\tiny S}}
\newcommand{\cl}{\mbox{\tiny CL}}
\newcommand{\BO}{\mathrm{BO}}

\newcommand{\closed}{\rm c}
\newcommand{\bo}{\bm{\Omega}}

\newcommand{\bp}{\tilde{\bm{\phi}}}
\newcommand{\bv}{\bm{V}}
\newcommand{\bj}{\bm{J}}
\newcommand{\cc}{\bm{\chi}}
\newcommand{\bi}{\mathbf{I}}
\newcommand{\cA}{\mathcal{A}}
\newcommand{\bc}{\bm{c}}
\newcommand{\bC}{\bm{C}}

\newcommand{\hP}{\hat{P}}
\newcommand{\hQ}{\hat{Q}}
\newcommand{\hp}{\hat{p}}
\newcommand{\hx}{\hat{x}}
\newcommand{\hF}{\hat{F}}

\newcommand{\as}{\alpha}
\newcommand{\nl}{\nonumber \\}
\newcommand{\ii}[3]{\int_{#1}^{#2}\mathrm{d}#3\,}
\newcommand{\der}[1]{\frac{\mathrm{d}}{\mathrm{d}#1}}
\newcommand{\eq}[1]{Eq.\,(\ref{#1})}
\newcommand{\Eq}[1]{Equation\,(\ref{#1})}
\newcommand{\Eqtwo}[2]{Equations\,(\ref{#1}) and (\ref{#2})}
\newcommand{\eqs}[1]{Eqs.\,(\ref{#1})}
\newcommand{\eqm}[1]{Eq.\,(#1)}
\newcommand{\eqtwo}[2]{Eqs.\,(\ref{#1}) and (\ref{#2})}
\newcommand{\cf}[1]{[cf.\,Eq.\,(\ref{#1})]}
\newcommand{\cfm}[1]{cf.\,Eq.\,({#1})}
\newcommand{\App}[1]{Appendix\,\ref{#1}}
\newcommand{\Fig}[1]{Fig.\,\,\ref{#1}}
\newcommand{\Sec}[1]{Sec.\,\,\ref{#1}}
\newcommand{\citeref}[1]{Ref.\,\cite{#1}}

\newcommand{\markred}[1]{{\color{black}#1}}
\newcommand{\markblue}[1]{{\color{black}#1}}

\begin{document}
\title{\markred{Quantum Counterpart of Equipartition Theorem in Quadratic Systems}}

\author{Xin-Hai Tong*}
\affiliation{Department of Physics, The University of Tokyo, 5-1-5 Kashiwanoha, Kashiwa-shi, Chiba 277-8574, Japan}
\email[]{xinhai@iis.u-tokyo.ac.jp}

\begin{abstract}
The equipartition theorem is a fundamental law of classical statistical physics, which states that every degree of freedom contributes $k_{B}T/2$ to the energy, where $T$ is the temperature and $k_{B}$ is the Boltzmann constant. Recent studies have revealed the existence of a quantum version of the equipartition theorem.
In the present work,
we focus on how to obtain the quantum counterpart of the generalized equipartition theorem for arbitrary quadratic systems in which the multimode Brownian ocillators interact with multiple reservoirs at the same temperature. An alternative method of deriving the energy of the system is also discussed and compared with the result of the the quantum version of the equipartition theorem, after which we conclude that the latter is more reasonable. Our results can be viewed as an indispensable generalization of rencent works on a quantum version of the equipartition theorem.
\end{abstract}


\maketitle

\section{Introduction}
One of the elegant principles of classical statistical physics is the equipartition theorem, which has numerous applications in various topics, such as thermodynamics \cite{Dattagupta2010, Campisi2007, Koide2011}, astrophysics \cite{Abreu2020, Abreu2020a, Barboza2015} and applied physics \cite{Ziel1973, Sarpeshkar1993, Matheny2013}. \markred{It is natural to consider the quantum version of the equipartition theorem, since quantum mechanics has been founded over a hundred years and so as quantum statistical mechanics.}

Recent years have seen much progress on this topic. A novel work \cite{Bialas2019} investigated the simplest quantum Brownian oscillator model to formulate the energy of the system in terms of the averge energy of a quantum oscillator in a harmonic well. \markred{Based on this work, more papers \cite{Ghosh_2023, Kaur2021, Kaur2022a, Bialas2018, Spiechowicz2018, Spiechowicz2019, Spiechowicz2021, Kaur2023} tried to study quantum counterparts of the equipartition theorem} in different versions of quadratic open quantum systems from various perspectives, including electrical circuits \cite{Ghosh_2023}, dissipative diamagnetism \cite{Kaur2023} \markblue{and focusing on kinetic energy for a more general setup \cite{luczka2020quantum}.}

In this work, we aim to deduce a \markred{quantum counterpart of the generalized equipartition theorem \cite{Magnano2020}} for arbitray open quantum quadratic systems. Many quadratic systems share the same algebra $[\hat{A},\hat{B}]=i\hbar$, where the 
binary operator pair can be the coordinate $\hat{x}$ and the momentum $\hat{p}$ for oscillators, the 
magnetic flux $\hat{\Phi}$ and charge $\hat{Q}$ for quantum circuits, and so on. We here turn to Brownian oscillators as an example to grasp the physical nature of all these systems. 

To construct such systems, we adopt a generalized Calderia-Leggett model \cite{Caldeira1981} and manage to transform it into a multi-mode Brownian-oscillator system well-discussed in \citeref{Tong2023}. For generality, we do not choose a concrete form of dissipation. We also generalize a formula in \citeref{Ford1988} for the internal energy so that it could be applied to the multi-mode Brownian oscillator system. It has been debated \cite{Ghosh2023} which of  the formula in \citeref{Ford1988} or the \markblue{quantum counterpart of the equipartition theorem} given in \citeref{Bialas2019} truly describes the energy of the system. Our analysis shows that the generalized version of the former formula cannot be used to find the energy, which implies that the latter one is more reasonable.

\markred{
The remainder of this paper is organized as follows. In \Sec{sec2}, we construct a quadratic system from the Calderia-Leggett model. In \Sec{sec3} we deduce the generalized equipartition theorem for this system and show its link to the conventioal equipartition theorem. In \Sec{sec4} we try to give a multi-mode version of the remarkable formula in \citeref{Ford1988}. More theoretical details are given in Appendixes. Numerical results are demonstrated in \Sec{sec5}.
 We summarize this paper in \Sec{sec6}.
Through out this paper, we set $\hbar=1$ and $\beta=1/(k_{\B}T)$ with $k_{\B}$ being the Boltzmann constant and $T$ being the temperature of the reservoirs if there is no special reminder.
}

\section{Arbitrary quadratic system}\label{sec2}
To construct our arbitrary quadratic system, let us start with the multi-mode Calderia-Leggett model \cite{Caldeira1981} 
\begin{align}\label{cl}
	H_{\cl}&=\sum_{u}\frac{\hP^{2}_{u}}{2M_{u}}+\sum_{uv}\frac{1}{2}k_{uv}\hQ_{u}\hQ_{v}+\sum_{\as j}\bigg[\frac{\hp^{2}_{\as j}}{2m_{\as j}}
	\nl &\quad\quad+\frac{1}{2}m_{\as j}\omega^{2}_{\as j}\bigg(\hx_{\as j}+\frac{1}{m_{\as j}\omega^{2}_{\as j}}\sum_{u}c_{\as uj}\hQ_{u} \bigg)^{2} \bigg],
\end{align}
where $u,v\in\{1,2,...,n_{\s}\}$ and $j\in\{1,2,...,n_{\as}\}$ are indices for the oscillators in the system and in the $\as$th bath,  repectively. The coefficient $c_{\as uj}$ represents the coupling strength between the coordinate of the $u$th oscillator in the system and the $j$th oscillator in the $\as$th bath. The convention would put $-c_{\as uj}$ in the last term of \eq{cl}, but we replace it by $c_{\as uj}$. We also have the commutation relations for all the momentum and position operators as follows:
\begin{align}
	[\hQ_{u},\hP_{v}]=i\delta_{uv},\quad[\hx_{\as_{1} j_{1}},\hp_{\as_{2} j_{2}}]=i\delta_{\as_{1}\as_{2}}\delta_{j_{1}j_{2}}
\end{align}
with $\delta$ representing the Kronecker delta.
Equation\,(\ref{cl}) can be reorganized in the following forms:
\begin{subequations}\label{HT}
	\begin{align}
		&H_{\cl}=H_{\s}+H_{\s\B}+h_{\B},\label{hcl}\\
		&H_{\s}=\sum_{u}\frac{\hP^{2}_{u}}{2M_{u}}+\frac{1}{2}\sum_{uv}\bigg(k_{uv}+\frac{c_{\as u j}c_{\as v j}}{m_{\as j}\omega^{2}_{\as j}}\bigg)\hQ_{u}\hQ_{v}\label{Hs1},\\
		&H_{\s\B}=\sum_{\as u} \hQ_{u}\hF_{\as u}\label{Hsb},\quad\hF_{\as u}=\sum_{j}c_{\as uj}\hx_{\as j},\\
		&h_{\B}=\sum_{\as}h_{\as\B}=\sum_{\as j}\bigg( \frac{\hp^{2}_{\as j}}{2m_{\as j}}+\frac{1}{2}m_{\as j}\omega^{2}_{\as j}\hx^{2}_{\as j}\bigg)\label{hb}.
	\end{align}
\end{subequations}
Here, the system-bath interaction results from the linear coupling of the system coordinate $\hQ_{u}$ and the random force $\hF_{\as u}$. 
We also emphasize that
all the mutually independent baths $\{h_{\as \B}\}$ in \eq{hb} are at the same inverse temperature $\beta$.
By defining the pure bath response function as
\begin{align}\label{phi}
	\bm{\phi}_{\as}(t)\equiv\{\phi_{\as uv}(t)\equiv i\langle [\hF^{\B}_{\as u}(t),\hF^{\B}_{\as v}(0)]  \rangle_{\B}\},
\end{align}
we recognize that
\begin{align}\label{5}
	\sum_{j}\frac{c_{\as u j}c_{\as v j}}{m_{\as j}\omega^{2}_{\as j}}=\tilde{\phi}_{\as uv}(0),
\end{align}
where $\hF^{\B}_{\as u}(t)\equiv e^{ih_{\B}t}\hF_{\as u}e^{-ih_{\B}t}$ and the average is defined over the canonical ensembles of baths as in $ \langle  \cdots \rangle_{\B}\equiv \text{tr}_{\B}[\cdots\,\otimes_{\alpha}e^{-\beta h_{\alpha\B}}]/\prod_{\as}\text{tr}_{\B}(e^{-\beta h_{\alpha\B}})$.
In \eq{5} we use tilde to denote the Laplace transform $\tilde{f}(\omega)=\ii{0}{\infty}{\omega}e^{i\omega t}f(t)$ for any function $f(t)$. 
By denoting 
$V_{uv}\equiv k_{uv}+\sum_{\as}\tilde{\phi}_{\as uv}(0)$ and $ \Omega_{u}\equiv M^{-1}_{u}$ for convenience,
we rewrite \eq{Hs1} in the form
\begin{align}\label{Hs}
	H_{\s}&=H_{\BO}+H_{\rm ren}
	\nl&\equiv\bigg[ \frac{1}{2}\sum_{u}\Omega_{u}\hat{P}^{2}_{u}+\frac{1}{2}\sum_{uv}\Big(V_{uv}-\sum_{\as}\tilde{\phi}_{\as uv}(0)\Big)\hat{Q}_{u}\hat{Q}_{v}\bigg]
	\nl &\quad+\bigg[
	\frac{1}{2}\sum_{\as uv}\tilde{\phi}_{\as uv}(0)\hQ_{u}\hQ_{v} \bigg],
\end{align}
which is the starting point of our \markred{quantum counterpart of the equipartition theorem.} Here, $H_{\rm ren}$ denotes the renormalization energy. The system Hamiltonian, referred to as \eq{Hs}, now is identical to the one presented in \citeref{Tong2023}. Physically, we need $\bm{V}=\{V_{uv}\}$, $\bm{k}=\{k_{uv}\}$ and $\bo=\{\Omega_{u}\delta_{uv}\}$ to be positive definite. Without loss of generality, we can always set $\bv$ and $\bm{k}$ to be symmetric. 
Detailed derivation of \eq{5} can be found in \App{appa}. 

\section{Quantum Counterpart of Generalizad equipartition theorem}\label{sec3}
The conventional \markred{quantum counterpart of the equipartition theorem for the single-mode Calderia-Leggett model deals with the kinetic energy $E_{\mathrm{k}}(\beta)$  in the Gibbs state of the total system with the inverse temperature being $\beta$ \cite{Bialas2019} in the form of
\begin{align}\label{7}
	E_{\mathrm{k}}(\beta)&=\mathbb{E}_{\mathrm{k}}\bigg[\frac{\hbar\omega}{4}\coth\frac{\beta\hbar\omega}{2}\bigg]\nl &\coloneq\ii{0}{\infty}{\omega}\mathbb{P}_{k}(\omega)\frac{\hbar\omega}{4}\coth\frac{\beta\hbar\omega}{2}.
\end{align}
Here, we temporarily add $\hbar$ for later convenience and $\mathbb{E}_{\mathrm{k}}[f(\omega)]$ denotes the expectation of a function $f(\omega)$ over the normalized distribution function $\mathbb{P}_{\mathrm{k}}(\omega)$, which satisifies $\mathbb{P}_{\mathrm{k}}(\omega)\geq 0$ and $\ii{0}{\infty}{\omega}\mathbb{P}_{\mathrm{k}}(\omega)=1$. \Eq{7} can be reduced to the classical case since $\lim_{\hbar\rightarrow 0}E_{\mathrm{k}}(\beta)=\mathbb{E}_{\mathrm{k}}[\lim_{\hbar\rightarrow 0}(\hbar\omega/4)\coth(\hbar\beta\omega/2)]=\mathbb{E}_{\mathrm{k}}[1/2\beta]=1/2\beta$.
However, } when some degrees of freedom are interwined with each other, such as in our model \cf{HT}, we would better use the generalized the equipartition theorem \cite{Magnano2020}. 

In the rest of this work we study the quantity $	\langle \hat{X}_{i}\partial H_{\BO}/\partial \hat{X}_{j}  \rangle$ for any system degrees of freedom $\hat{X}_{i},\hat{X}_{j}\in\{\hP_{u}\}\cup\{\hQ_{u}\}$, \markred{with the average defined in the total Gibbs state  $\langle...\rangle\coloneq\operatorname{tr}_{\T}[...e^{-\beta H_{\cl}}]/\operatorname{tr}_{\T}e^{-\beta H_{\cl}}$, which is well defined since we assume that all the bath are at the same inverse temperature $\beta$. The derivative of the operator here is merely a notation, indicating that we initially treat all distinct operators as mutually independent variables like real numbers. After obtaining the result, We restore these variables back to operators.} In the main text we focus on the cases  $\hat{X}_i=\hat{X}_j$, while other cases are discussed in \App{appc}.

We have $	\langle  \hat{Q}_{u}\partial H_{\BO}/\partial \hQ_{u} \rangle 
=\sum_{v}[V_{uv}-\sum_{\as}\tilde{\phi}_{\as uv}(0)]\langle \hat{Q}_{u}\hat{Q}_{v}  \rangle$ when we choose $\hat{X}_{i}=\hat{X}_{j}=\hQ_{u}$.
% whose partial trace of bath is the system
%equilibrium state: $\rho_{\s}:=\operatorname{tr}_{\B}\rho^{\rm st}_{\T}=e^{-\beta H_{\s}}/\operatorname{tr}_{\s}e^{-\beta H_{\s}}$.
Under the help of the fluctuation-disspation theorem \cite{Callen1951} for symmetrized correlation function \cite{Ghosh2023}, we obtain
\begin{align}\label{Q}
	\frac{1}{2}\bigg\langle  \qty{\hat{Q}_{u},\frac{\partial H_{\BO}}{\partial \hat{Q}_{u}}} \bigg\rangle
	 &=\sum_{v}\frac{1}{\pi}\ii{0}{\infty}{\omega}\bigg[V_{uv}
	 \nl &\quad-\sum_{\as}\tilde{\phi}_{\as uv}(0)\bigg]\operatorname{Re}J^{QQ}_{uv}(\omega)\coth\frac{\beta\omega}{2},
\end{align}
with $\bj^{QQ}(\omega)=\{J^{QQ}_{uv}(\omega)\}$ being the anti-Hermitian part of the matrix $\tilde{\cc}^{QQ}(\omega)=\{\tilde{\chi}^{QQ}_{uv}(\omega)\}$ and $\qty{ \bullet, \circ  }$ representing the anticommutator.
%($\cc^{QQ}(\omega)$ is symmetric since we can always set $\bv$ be symmetric, so anti--Hermitian part here is equivalent to imaginary part):
Here, we denote the system response function of any two operators $\hat{A}_{u}$ and $\hat{B}_{v}$ as $\chi^{AB}_{uv}(t)\equiv i\langle [\hat{A}_{u}(t),\hat{B}_{v}(0)]\rangle$. \markred{According to \citeref{Tong2023},  we also have some useful relations for the quantities like $\tilde{\cc}^{AB}(\omega)$. We list them below for later convenience:
\begin{subequations}
	\begin{align}
			&\tilde{\cc}^{QQ}(\omega)=\Big
		[\bo\bv-\omega^2 \bi-\bo \sum_\alpha \bp_\alpha(\omega)\Big]^{-1} \bo ,\label{cc}\\
	&	\tilde{\cc}^{PP}(\omega)=\bo^{-1}+\omega^{2}\bo^{-1}\tilde{\cc}^{QQ}(\omega)\bo^{-1},\label{chiPP}\\
	&\tilde{\cc}^{QP}(\omega)=i\omega\tilde{\cc}^{QQ}(\omega)\bo^{-1},\label{chiqp}
	\\
	&
	\tilde{\cc}^{PQ}(\omega)=-i\omega\bo^{-1}\tilde{\cc}^{QQ}(\omega)\label{chipq}
	,
	\end{align}
\end{subequations}
}where $\bv$ and $\bo$ are given below \eq{Hs}. Equation\,(\ref{Q}) can be recast as
\markred{
\begin{align}\label{Qeet}
	\bigg\langle  \hat{Q}_{u}\frac{\partial H_{\BO}}{\partial \hat{Q}_{u}} \bigg\rangle&=\ii{0}{\infty}{\omega}\mathbb{P}_{Q_uQ_u}(\omega)\frac{\omega}{2}\coth\frac{\beta\omega}{2}
\end{align}
with
\begin{align}\label{PQ}
	\mathbb{P}_{Q_uQ_u}(\omega)=\frac{2}{\pi\omega}\sum_{v}\bigg[V_{uv}-\sum_{\as}\tilde{\phi}_{\as uv}(0)\bigg]J^{QQ}_{uv}(\omega),
\end{align}
}where we used the fact that $\bj^{QQ}(\omega)$ is symmetric since $\tilde{\cc}^{QQ}(\omega)$ is symmetric and therefore $\bj^{QQ}\qty(\omega)$ is real. 

A similar process for the case $\hat{X}_{i}=\hat{X}_{j}=\hP_{u}$ leads to $	\langle  \hP_{u}\partial H_{\BO}/\partial \hP_{u} \rangle 
=\Omega_{u}\langle \hat{P}^{2}_{u}  \rangle$. Using the fluctuation-dissipation theorem \cite{Callen1951} again, we find
\begin{align}\label{P1}
	\frac{1}{2}\bigg\langle  \qty{\hat{P}_{u},\frac{\partial H_{\BO}}{\partial \hat{P}_{u}}} \bigg\rangle
	&=\frac{\Omega_{u}}{\pi}\ii{0}{\infty}{\omega}\operatorname{Re}J^{PP}_{uu}(\omega)\coth\frac{\beta\omega}{2},
\end{align}
where
 $\bj^{PP}(\omega)=\{J^{PP}_{uv}(\omega)\}$ is the 
anti-Hermitian part of matrix $\tilde{\cc}^{PP}(\omega)=\{\tilde{\chi}^{PP}_{uv}(\omega)\}$ \cf{chiPP}.
\begin{comment}
We obtain
\begin{align}\label{Pc}
	\bigg\langle \hP_{u}\frac{\partial H_{\BO}}{\partial \hP_{u}}  \bigg\rangle_{\closed}(\omega,\beta)=\frac{\omega}{2}\coth\frac{\beta\omega}{2},
\end{align}
for the same closed system with $c_{\as uj}=0$ and $\bv=\omega^{2}\bo^{-1}$. It is evident that the left-hand side of \eq{Pc} equals twice the  average kinetic energy of a harmonic oscillator in canonical ensemble.	
\end{comment}
\markred{
By substituting \eq{chiPP} into \eq{P1}, we obtain the final result
\begin{align}\label{Peet}
	\bigg\langle  \hat{P}_{u}\frac{\partial H_{\BO}}{\partial \hat{P}_{u}} \bigg\rangle=\ii{0}{\infty}{\omega}\mathbb{P}_{P_{u}P_{u}}(\omega)\frac{\omega}{2}\coth\frac{\beta\omega}{2}
\end{align}
with
\begin{align}\label{PP}
	\mathbb{P}_{P_uP_u}(\omega)=\frac{2\omega}{\pi\Omega_{u}}J^{QQ}_{uu}(\omega).
\end{align}

The proof of positivity and normalization of \eqtwo{PQ}{PP} can be found in \App{appb}, by which we can recast \eqtwo{Qeet}{Peet} as
\begin{align}\label{16}
\bigg\langle  \hat{X}_{i}\frac{\partial H_{\BO}}{\partial \hat{X}_{i}} \bigg\rangle&=\mathbb{E}_{ii}\bigg[ \frac{\omega}{2}\coth\frac{\beta\omega}{2} \bigg]
	\nl &\coloneq\ii{0}{\infty}{\omega}\mathbb{P}_{X_iX_i}(\omega)\frac{\omega}{2}\coth\frac{\beta\omega}{2}
\end{align}
with
\begin{align}\label{171}
	\mathbb{P}_{ii}(\omega)\geq 0\quad\text{and}\quad \ii{0}{\infty}{\omega}\mathbb{P}_{ii}(\omega)=1,
\end{align}
where we use the notation $\mathbb{P}_{ii}$ rather than $\mathbb{P}_{X_{i}X_{i}}$ for simplicity. 
In \App{appc}, we extend \eqtwo{16}{171} to the cases where $\hat{X}_i\neq\hat{X}_j$ and summarize all the results as 
\begin{align}\label{18}
\frac{1}{2}	\left\langle \bigg\{ \hat{X}_{i},\frac{\partial H_{\BO}}{\partial \hat{X}_{j}} \bigg\}\right\rangle&=\mathbb{E}_{ij}\bigg[ \frac{\hbar\omega}{2}\coth\frac{\hbar\beta\omega}{2} \bigg]
\end{align}
for any system degrees of freedom $\hat{X}_{i},\hat{X}_{j}\in\{\hP_{u}\}\cup\{\hQ_{u}\}$
with
\begin{align}\label{19}
	\mathbb{P}_{ii}(\omega)\geq 0\quad\text{and}\quad \ii{0}{\infty}{\omega}\mathbb{P}_{ij}(\omega)=\delta_{ij}
\end{align}
and
$\mathbb{E}_{ij}[...]$ denotes the expectation over $\mathbb{P}_{ij}(\omega)$.
Here, we temporarily add $\hbar$ for later convenience and let $\delta$ denote the Kronecker delta. \Eqtwo{18}{19} are partly the main results of the present work.}

Discussions are presented here to conclude this section. Once we take the classical limit $\hbar\rightarrow 0$ and the weak-coupling limit $c_{\as uj}\rightarrow 0$, \eq{18} reduces to $ \langle X_{i}\partial H_{\s}/\partial X_{j}  \rangle=\delta_{ij}/\beta$ \cf{Hs}, which is termed as the generalized equipartition theorem  \cite{Magnano2020}. We also emphasize that though the right-hand side of \eq{18} depends on different degrees of freedom ($i$ and $j$), the function $(\omega/2)\coth(\beta\omega/2)$ is universal for all the degrees of freedom, which is the ``equipartition" in a quantum sense. Therefore, \eq{18} is termed as the quantum counterpart of the generalized equipartition theorem. It is evident that \eqtwo{PQ}{PP} reduce to the results in Refs\,.\cite{Bialas2019,Ghosh2023} for the single mode $n_{\s}=1$ case. \markblue{Besides, by noticing that $\langle  \hP_{u}\partial H_{\BO}/\partial \hP_{u} \rangle$ equals twice the kinetic energy of the $u$th oscillator and $\bj^{PP}(\omega)=\omega\bo^{-1}\bj^{QQ}(\omega)\bo^{-1}$ \cf{chiPP}, \eqtwo{Peet}{PP} reduce to the results presented in \citeref{luczka2020quantum}.
}

\markred{\Eq{18} here offers a new angle on how to calculate the quantities of 
open systems, which is generally hard to obtain.
An application is given below.} Noting that the total energy is given by \markred{$E(\beta)=1/2\sum_{i}\langle  \hat{X}_{i}\partial H_{\BO}/\partial \hat{X}_{i} \rangle
$, we arrvie at
\begin{align}\label{E}
	E(\beta)= \mathbb{E}\bigg[\frac{n_{\s}\omega}{2}\coth\frac{\beta\omega}{2} \bigg]
\end{align}
with $\mathbb{E}[...]$ denotes the expectation over
\begin{align}\label{P}
	\mathbb{P}(\omega)\coloneq \sum_{i}\mathbb{P}_{ii}(\omega)/2n_{\s},
\end{align}
which is checked \cf{171} to be nonnegative and normalized over $\mathbb{R}^{+}$.}
\Eq{E} is termed as the \markred{quantum counterpart of conventional equipartition theorem \cite{Ghosh2023}.} Moving further with the help of thermodynamic equations, we can determine the free energy $F(\beta)$ of the system by
\begin{align}\label{th-F}
	F(\beta)+\beta\frac{\partial F(\beta)}{\partial \beta}=E(\beta),
\end{align}
and hence \eqtwo{th-F}{E} may yield \markred{
\begin{align}\label{F}
	F(\beta)=\mathbb{E}\bigg[ \frac{n_{\s}}{\beta}\ln\bigg(2\sinh\frac{\beta\omega}{2}\bigg)\bigg],
\end{align}
from which we further obtain an expression for the partition function of the system in the form of 
\begin{align}\label{Zs}
	\ln Z_{\s}(\beta)=-\mathbb{E}\bigg[ n_{\s}\ln\bigg(2\sinh\frac{\beta\omega}{2}\bigg)\bigg].
\end{align}
}
Note that \eq{Zs} is much easier to obtain than the conventional 
influence-functional approach \cite{Feynman2000, Ingold2002}.

\section{Alternative approach for the energy}\label{sec4}
A recent review \cite{Ghosh2023} presented another approach to obtain the energy of the system of multi-mode harmonic oscillators. When introduced in \citeref{Ford1988} first, the result was only limited to the single-mode case. Here we generalize their derivation and find that their derivation is not applicable to the multi-mode case.

The starting point of \citeref{Ford1988} is quite straightforward. Since the conventional definition for the internal energy of the oscillator \markred{ $U_{\s}(\beta)=\operatorname{tr}_{\T}[H_{\BO}e^{-\beta H_{\cl}}]/\operatorname{tr}e^{-\beta H_{\cl}}$} is generally challenging to handle, we adopt a normal-mode coordinates, so that the transformed Hamiltonian $H_{\T}$ describes $N(=1+\sum_{\as}n_{\as})$ uncoupled oscillators. Physically we do not need to obtain the detailed information for any normal modes, since the total energy $U_{\T}(\beta)$ is only associated with  the normal frequencies, namely
\begin{align}\label{UT}
	U_{\T}(\beta)=\sum_{r=1}^{N}u(\overline{\omega}_{r},\beta)\coloneq\sum_{r=1}^{N}\frac{\overline{\omega}_{r}}{2}\coth\frac{\beta\overline{\omega}_{r}}{2},
\end{align}
with $\overline{\omega}_{r}$ being the normal frequency for $r$-th oscillator in the transformed system. Here, we also introduced the notation $u(\omega,\beta)\equiv(\omega/2)\coth(\beta\omega/2)$ for later convenience. Since the energy for the independent bath is well-defined as
\begin{align}\label{UB}
	U_{\B}(\beta)=\sum_{\as j}\frac{\omega_{\as j}}{2}\coth\frac{\beta\omega_{\as j}}{2},
\end{align}
the authors of \citeref{Ford1988} interpreted the difference 
\begin{align}\label{Us}
	U_{\s}(\beta)=U_{\T}(\beta)-U_{\B}(\beta)
\end{align}
as the internal energy and found it to be
\markred{
\begin{align}
	U_{\s}(\beta)=\ii{0}{\infty}{\omega}\frac{1}{\pi }\operatorname{Im}\der{\omega}\ln\tilde{\chi}^{QQ}(\omega)\frac{\omega}{2}\coth\frac{\beta\omega}{2},
\end{align}
where $\tilde{\chi}^{QQ}(\omega)$ is the one-dimensional version of \eq{cc}.

Following their procedures for the multi-mode case, we find (see \App{appd} for detailed derivation)
\begin{align}\label{final}
	U_{\T}(\beta)-n_{\s}U_{\B}(\beta)=\ii{0}{\infty}{\omega}\mathbb{B}(\omega)n_{\s}\frac{\omega}{2}\coth\frac{\beta\omega}{2}
\end{align} 
with
\begin{align}\label{B}
	\mathbb{B}(\omega)=\frac{1}{\pi n_{\s}}\operatorname{Im}\der{\omega}\ln\det\tilde{\cc}^{QQ}(\omega),
\end{align}
}which does not give us $U_{\s}(\beta)=U_{\T}(\beta)-U_{\B}(\beta)$. On the other hand, the result of the system internal energy according to \markred{the quantum counterpart of equipartition theorem \cf{E}} is applicable to any multi-mode case. Therefore, we conclude that \eq{E} is more reasonable than the alternative approach dicussed in \citeref{Ford1988} as an expression for the internal energy of the system.
\markred{
	\Eqtwo{final}{B} are another part of the main results of this work.}

\section{Numerical demonstrations}\label{sec5}
\markred{
In this section, we use the two-mode ($u,v\in\{1,2\}$) Brownian-oscillator system coupled with one reservoir ($\as=1$) to give a numerical demonstration of our results. The system Hamiltonian $H_{\s}$ of the two-mode system reads $H_{\s}=(\Omega_{1}\hP^{2}_{1}+\Omega_{2}\hP^{2}_{2})/2+(V_{11}\hQ^{2}_{1}+V_{22}\hQ^{2}_{2}+2V_{12}\hQ_{1}\hQ_{2})/2$, while the system-bath interaction term becomes $H_{\s\B}=\hQ_{1}\hF_{1}+\hQ_{2}\hF_{2}$ with the random force $\hF_{u1}=\sum_{j}c_{uj}\hx_{j}$ for $u\in\{1,2\}$. The bath Hamiltonian reduces to $h_{\B}=\sum_{ j}( \hp^{2}_{ j}/2m_{j}+m_{ j}\omega^{2}_{j}\hx^{2}_{j}/2)$. 
}
To enhance clarity, we choose the spectrum of the pure bath in the following form:
\begin{align}\label{BO}
	\bp(\omega,\lambda)=\lambda\bm{\eta}\operatorname{Im} \frac{\Omega^{2}_{\B}}{\Omega^{2}_{\B}-\omega^{2}-i\omega\gamma_{\B}}
\end{align}
with $\bm{\eta}\equiv\{\eta_{uv}=\eta_{u}\delta_{uv}\}$ specifying the strength of the system-bath couplings. \markred{From an experimental point of view, this setup can be realized, for example, in molecular junctions \cite{giazotto2006opportunities,saira2007heat,bell2008cooling}.} We also introduce the parameter $\lambda\in\{1,1.25,1.5\}$ to vary the strength, \markred{which can be realized experimentally by modifying the intermolecular distance. Through out this section, we select the parameters in the unit of $\Omega_{\B}$ as $\gamma_{\B}=1.25\Omega_{\B}$, $\Omega_{1}=\Omega_{2}=V_{11}=V_{22}=\Omega_{\B}$ and $V_{12}=V_{21}=0.5\Omega_{\B}$. 
The strength of the couplings are chosen to be $\eta_{1}=0.2\Omega_{\B}$ and $ \eta_{2}=0.1\Omega_{\B}$.}

\begin{figure}[h]
	\centering
	\begin{subfigure}[b]{0.45\textwidth}
		\includegraphics[width=\textwidth]{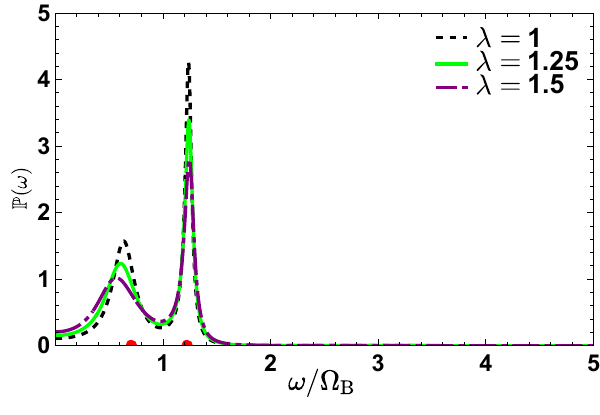}
		\caption{Plot of $\mathbb{P}(\omega)$ for $\lambda\markred{\in\{1,1.25,1.5\}}$}
		\label{1}
	\end{subfigure}
	\begin{subfigure}[b]{0.45\textwidth}
		\includegraphics[width=\textwidth]{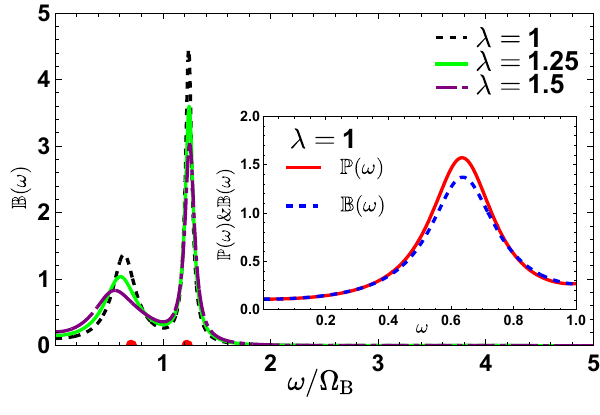}
		\caption{Plot of $\mathbb{B}(\omega)$ for $\lambda\markred{\in\{1,1.25,1.5\}}$. The subgraph compares $\mathbb{P}(\omega)$ with $\mathbb{B}(\omega)$ in the case $\lambda=1$. }
		\label{2}
	\end{subfigure}
	\captionsetup{justification=raggedright,singlelinecheck=false}
	\caption{Plots of $\mathbb{P}(\omega)$ in (a) and $\mathbb{B}(\omega)$ in (b) when $\lambda\in\{1,1.25,1.5\}$. Here the red dots on the horizontal axes represent \markred{ $\omega=0.7071\Omega_{\B}$ and $\omega=1.224\Omega_{\B}$,} respectively, which are the square root of eigenvalues of the matrix $\bo\bv$ according to the parameters chosen in the main text.}
	\label{11}
\end{figure}

Figures (\ref{1}) and (\ref{2}) depict $\mathbb{P}(\omega)$ and $\mathbb{B}(\omega)$ in the three cases.
\begin{comment}
	It is evident that  the square root of  the eigenvalues of $\bo\bv$ correspond to the maximum points of $\mathbb{P}(\omega)$ and $\mathbb{B}(\omega)$, giving the most probable frequences of these two distributions. 	
\end{comment}
As $\lambda$ decreases, the curves become sharper around the square root of  the eigenvalues of $\bo\bv$. In other words, the maximum points of $\mathbb{P}(\omega)$ and $\mathbb{B}(\omega)$ become closer to them as $\lambda$ decreases.
In fact, we can prove the following (see \App{appe})
\markred{
\begin{align}\label{301}
	\lim\limits_{\{c_{\as uj}\}\rightarrow \{0\}}\mathbb{P}(\omega)=\frac{1}{n_{\s}}\sum_{\omega^{2}_{i}\in \operatorname{spec}(\bo\bv)}\big[\delta(\omega-\omega_{i}) +\delta(\omega+\omega_{i})\big],
\end{align}}where the summation is over all the square root of eigenvalues of the matrix $\bo\bv$ (considering multiplicity). These results show that in the weak-coupling limit, \markred{only the oscillators with these typical frequencies} contribute to the quantity that we consider, such as the energy. 
\markred{In this case, the energy reads
\begin{align}\label{39}
	E_{\mathrm{weak}}(\beta)=\sum_{\omega^{2}_{i}\in \operatorname{spec}(\bo\bv)}\frac{\omega_{i}}{2}\coth\frac{\beta\omega_{i}}{2},
\end{align} 
which is how the quantum counterpart of equipartition theorem behaves in the weak-coupling limit. In the single-mode case, a similar pattern has also been discussed in \citeref{Ghosh2023}
}
\section{Summary}\label{sec6}
To summarize, we derived \markred{a quantum counterpart of the generalized equipartition for arbitray quadratic systems,} \markblue{which we can also reduce to the results presented in previous works for the single-mode case.} We also extended another formula for the internal energy of the multi-mode Brownian-oscillator system. The generalized formula as well as our analysis shed light on the controversies upon the method. We noticed that our \markred{quantum counterpart of equipartition theorem} can be used to obtain the partition function of the system in a much easier way than the classical approach \cite{Ingold2002}. Our results can be viewed as an indispensable \markred{development of rencent works on the quantum counterpart of the  equipartition theorem.} 

As a future prospect, expressing thermodynamic quantities as an infinite series also offers potential advantages for this objective \cite{Ghosh2023}. Work in this direction is in progress. As another point, it seems difficult to discuss the quantum version of equipartition theorem without the help of fluctuation-disspation theorem and to consider it over steady states or even in general nonequilibrium. Discussing the present topic under other more difficult setups, such as quartic systems, is also challenging. All of them constitute directions of further
development.
\section*{Acknowledgements}
The present author thanks Prof. Naomichi Hatano for his valuable suggestions, as well as to Prof. Zongping Gong and Ao Yuan for their fruitful discussions.
Xinhai Tong was supported by FoPM, WINGS Program, the University of Tokyo. This research was supported by Forefront Physics and Mathematics Program to Drive Transformation (FoPM), a World-leading Innovative Graduate Study (WINGS) Program, the University of Tokyo.

\appendix
\section{Derivation of \eq{5}}\label{appa}
Here we show the detailed derivation of \eq{5}. Let us start from the Heisenberg equation of motion for any operator, $\dot{\hat{X}}(t)=i[H_{\T},\hat{X}(t)]$. Focusing on the bath quantities $\hx_{\as j}$ and $\hp_{\as j}$, we obtain
\begin{subequations}
	\begin{align}
		&\dot{\hx}_{\as j}(t)=\hp_{\as j}(t)/m_{\as j},\label{A1a}\\
		&\dot{\hp}_{\as j}(t)=-m_{\as j}\omega^{2}_{\as j}\hx_{\as j}(t)-\sum_{u}c_{\as uj}\hQ_{u}(t)\label{A1b}.
	\end{align}
\end{subequations}
Taking the time deriative of \eq{A1a} and put \eq{A1b} into it, we obtain the following equation:
\begin{align}\label{ddotx}
	\ddot{\hx}_{\as j}(t)=-\omega^{2}_{\as  j}\hx_{\as j}(t)-\sum_{u}\frac{c_{\as uj}}{m_{\as j}}\hQ_{u}(t).
\end{align}
One can verify the formal solution to \eq{ddotx} is
\begin{align}
	\hx_{\as j}(t)&=\hx_{\as j}(0)\cos \omega_{\as j}t+\frac{\hp_{\as j}(0)}{m_{\as j}\omega_{\as j}}\sin\omega_{\as j}t
	\nl &\quad+\sum_{u}\frac{c_{\as uj}}{m_{\as j}\omega_{\as j}}\ii{0}{t}{\tau}\sin\omega_{\as j}(t-\tau)\hQ_{u}(\tau).
\end{align}
Meanwhile we know from the definition of $\hF^{\B}_{\as u}$ that
\begin{align}\label{FB}
	\hF^{\B}_{\as u}(t)=\sum_{j}c_{\as uj}\bigg[\hx_{\as j}(0)\cos\omega_{\as j}t+\frac{\hp_{\as j}(0)}{m_{\as j}\omega_{\as j}}\sin\omega_{\as j}t \bigg]
\end{align}
Putting \eq{FB} into \eq{phi}, we directly obtain
\begin{align}\label{A5}
	&\phi_{\as uv}(t)=\sum_{j}\frac{c_{\as uj}c_{\as vj}}{m_{\as j}\omega_{\as j}}\sin\omega_{\as j}t 
	\nl &\Rightarrow
	\tilde{\phi}_{\as uv}(\omega)=\sum_{j}\frac{c_{\as uj}c_{\as vj}}{m_{\as j}\omega_{\as j}}\frac{\omega_{\as j}}{-\omega^{2}+\omega^{2}_{\as j}}.
\end{align}
Thus obviously we find \eq{5}. We can also see that $\bp_{\as}(\omega)$ is a symmetric matrix.

\section{Proof of Positivity and Normalization}\label{appb}
From the positivity of $\bj^{QQ}(\omega)$ \cite{Yan2005}, we directly know $\mathbb{P}_{P_uP_u}(\omega)\geq 0$. 
Once we rewrite \eq{PQ} as $\mathbb{P}_{Q_uQ_u}(\omega)=2\big[(\bv-\sum_{\as}\bp_{\as}(0))\bj^{QQ}(\omega)\big]_{uu}/\pi\omega$ (note that $\bj^{QQ}(\omega)$ is symmetric), its positivity becomes evident, since $\bm{k}\equiv\bv-\sum_{\as}\bp_{\as}(0)$ and $\bj^{QQ}(\omega)$ is also positive definite \cite{Yan2005}.
For the normalization of $\mathbb{P}_{Q_uQ_v}(\omega)$, we notice the following relation [\cfm{2.17} in \citeref{Yan2005}]: 
\begin{align}\label{B1}
	\tilde{\chi}^{QQ}_{uv}(0)=\frac{1}{\pi}\ii{-\infty}{\infty}{\omega}\frac{J^{QQ}_{uv}(\omega)}{\omega}.
\end{align}
One can notice that $J^{QQ}_{uv}(\omega)$ is an odd function, since $J_{uv}^{QQ}(-\omega)=-J^{QQ}_{vu}(\omega)$ \cite{Yan2005} and $\bj^{QQ}(\omega)$ is symmetric. We further obtain from \eq{B1} that
\begin{align}\label{B2}
	\big[(\bv-\sum_{\as}\bp_{\as}(0))^{-1}\big]_{uv}=\tilde{\chi}^{QQ}_{uv}(0)=\frac{2}{\pi}\ii{0}{\infty}{s}\frac{J^{QQ}_{uv}(\omega)}{\omega},
\end{align}
where the first equality comes from \eq{cc}.
\Eq{B2} is equivalent to
\begin{align}\label{B3}
\delta_{uw}=\sum_{v}\frac{2}{\pi}\int_{0}^{\infty}\mathrm{d}\omega\,\frac{J^{QQ}_{uv}(\omega)}{\omega} \big[\bv-\sum_{\alpha}\bp_{\alpha}(0)\big]_{vw}.
\end{align} 
In the special case of $u=w$ and considering the symmetry of  $\bv$,$\bp_{\as}(0)$ and $\tilde{\cc}^{QQ}(\omega)$, we can deduce that $\ii{0}{\infty}{\omega}\mathbb{P}_{Q_uQ_u}(\omega)=1.$

As for $\mathbb{P}_{P_uP_u}(\omega)$, 
from \eqm{2.17} in \citeref{Yan2005} we obtain
\begin{align}\label{B4}
	\dot{\chi}^{QQ}_{uv}(0)=\frac{1}{\pi}\ii{-\infty}{\infty}{\omega}\omega J^{QQ}_{uv}(\omega).
\end{align}
We can also know from \citeref{Tong2023} that $\dot{\cc}^{QQ}(t)=-\bo\cc^{QP}(t)$. Letting $t=0$, we obtain $\dot{\cc}^{QQ}(0)=\bo$ since $\chi^{QP}_{uv}(0)=i\langle [\hQ_{u},\hP_{v}]  \rangle=\delta_{uv}$. Using this result and \eq{B4} we arrive at $\ii{-\infty}{\infty}{\omega}\omega J^{QQ}_{uv}(\omega)=\pi\Omega_{u}\delta_{uv}$. By observing that $J^{QQ}_{uv}(\omega)$ is an odd function, we can further deduce the following:
\begin{align}\label{B5}
	2\ii{0}{\infty}{\omega}\omega J^{QQ}_{uv}(\omega)=\pi\Omega_{u}\delta_{uv},
\end{align}
which in the case of $u=v$ is equivalent to $ \ii{0}{\infty}{\omega}\mathbb{P}_{P_uP_u}(\omega)=1,$ and
thus we finish the proof for the normalization of $\mathbb{P}_{P_uP_u}(\omega)$.

\section{Analysis of the Cases $\hat{X}_i \neq \hat{X}_j$}\label{appc}
In this section we complete the derivation of our quantum counterpart of the generalized equipartition theorem by considering the cases of $\hat{X}_{i}\neq \hat{X}_{j}$. We start from 
\begin{align}\label{C1a}
		\frac{1}{2}\left\langle \qty{\hat{P}_{u},\frac{\partial H_{\BO}}{\partial\hat{P}_{v}}}\right\rangle & =\frac{1}{2}\Omega_{v}\left\langle \qty{\hat{P}_{u},\hat{P}_{v}}\right\rangle 
		\nl &=\frac{1}{2}\Omega_{v}\frac{2}{\pi}\ii{0}{\infty}{\omega}\operatorname{Re}J^{PP}_{uv}\qty(\omega)\coth\frac{\beta\omega}{2}
		  \nl
		&=\frac1{\pi\Omega_{u}}\int_{0}^{\infty}\mathrm{d}\omega\,\omega^{2}J^{QQ}_{uv}(\omega)\coth\frac{\beta\omega}2
		\nl &
		=\int_{0}^{\infty}\mathrm{d}\omega\,\mathbb{P}_{P_{u}P_{v}}(\omega)\frac{\omega}{2}\coth\frac{\beta\omega}{2}
\end{align}
with 
\begin{align}\label{C2}
	\mathbb{P}_{P_uP_v}(\omega)=\frac{2\omega}{\pi\Omega_{u}} J^{QQ}_{uv}(\omega),
\end{align}
where in the  third line we used the fact that $\bj^{PP}(\omega)=\{J^{PP}_{uv}(\omega)\}$ and $\bj^{QQ}(\omega)=\{J^{QQ}_{uv}(\omega)\}$ are the anti-Hermitian parts of \eqtwo{chiPP}{cc}, respectively. Using \eq{B5} we obtain
\begin{align}\label{C3}
	\ii{0}{\infty}{\omega}\mathbb{P}_{P_uP_v}(\omega)=\delta_{uv}.
\end{align}
Similarily, we evaluate 
\begin{align}\label{C4}
	&\quad\frac{1}{2}\left\langle \qty{\hat{Q}_{u},\frac{\partial H_{\BO}}{\partial\hat{Q}_{v}}}\right\rangle
	\nl &=\frac{1}{2}\sum_{w}\big[V_{vw}-\sum_{\alpha}\tilde{\phi}_{\alpha vw}(\omega)\big]\left\langle \qty{\hat{Q}_{u},\hat{Q}_{w}}\right\rangle
	\nl &=\frac{1}{2}\sum_{w}\frac{2}{\pi}\int_{0}^{\infty}\mathrm{d}\omega\, \bigg[V_{vw}
	-\sum_{\alpha}\tilde{\phi}_{\alpha vw}(\omega)\bigg]\operatorname{Re}J^{QQ}_{uw}(\omega)\coth\frac{\beta\omega}2
	\nl 
	&=\ii{0}{\infty}{\omega}\mathbb{P}_{Q_{u}Q_{v}}(\omega)\frac{\omega}{2}\coth\frac{\beta\omega}{2}
\end{align}
with
\begin{align}
	\mathbb{P}_{Q_uQ_v}(\omega)=\frac{2}{\pi\omega}\sum_{w}[V_{vw}-\sum_{\as}\tilde{\phi}_{\as vw}(\omega)]J^{QQ}_{uw}(\omega).
\end{align}
By utilizing \eq{B3}, we can derive the following result:
\begin{align}\label{C6}
	\ii{0}{\infty}{\omega}\mathbb{P}_{Q_uQ_v}(\omega)=\delta_{uv}.
\end{align}
As an additional outcome, we can also express  \eqtwo{C2}{C6} in the following alternative forms:
$\mathbb{P}_{P_{u}P_{v}}(\omega)=2\omega [\bj^{QQ}(\omega)\dot{\cc}^{-1}(0)]_{uv}/\pi$
 and $\mathbb{P}_{Q_{u}Q_{v}}(\omega)=2[\bj^{QQ}(\omega)\cc^{-1}(0)]_{uv}/\omega\pi$, respectively. 

Then we consider
\begin{align}\label{C7}
	\frac{1}{2}\left\langle \qty{\hat{Q}_{u},\frac{\partial H_{\BO}}{\partial\hat{P}_{v}}}\right\rangle&=\frac{1}{2}\Omega_{v}\left\langle \qty{\hat{Q}_{u},\hat{P}_{v}}\right\rangle
	\nl &=\frac{\Omega_{v}}{\pi}\ii{0}{\infty}{\omega}\operatorname{Re}J^{QP}_{uv}(\omega)\coth\frac{\beta\omega}{2}
	\nl &=\ii{0}{\infty}{\omega}\mathbb{P}_{Q_{u}P_{v}}(\omega)\frac{\omega}{2}\coth\frac{\beta\omega}{2}
\end{align}
with
\begin{align}
	\mathbb{P}_{Q_uP_v}(\omega)=\frac{2\Omega_{v}}{\pi\omega}\operatorname{Re}J^{QP}_{uv}(\omega),                       
\end{align}
where $\bj^{QP}(\omega)=\{J^{QP}_{uv}(\omega)\}$ is the anti-Hermitian part of the matrix $\tilde{\cc}^{QP}(\omega)=\{\tilde{\chi}^{QP}_{uv}(\omega)=\ii{0}{\infty}{\omega}e^{i\omega t}\chi^{QP}_{uv}(t)\}$. By \eq{chiqp}, we derive 
$
	\bj^{QP}(\omega)=\omega[\tilde{\cc}^{QQ}(\omega)\bo^{-1}+\bo^{-1}\tilde{\cc}^{QQ\dagger}(\omega)]/2,
$                            
which 
helps us to deduce 
\begin{align}\label{C9}
	\ii{0}{\infty}{\omega}\mathbb{P}_{Q_uP_v}(\omega)=0.
\end{align}
Here, we utilized the property that $\tilde{\cc}^{QQ}(\omega)$ is an even function in order to establish the following useful equality:
 $	\ii{0}{\infty}{\omega}\tilde{\cc}^{QQ}(\omega)=	\ii{-\infty}{\infty}{\omega}\tilde{\cc}^{QQ}(\omega)/2= \cc^{QQ}(0)=\bm{0},
$
which toghther with the expression for $\bj^{QP}(\omega)$ helps us to obtain \eq{C9}.
Following the similar process, we deduce
\begin{align}\label{C10}
	&\quad\frac{1}{2}\left\langle \qty{\hat{P}_{u},\frac{\partial H_{\BO}}{\partial\hat{Q}_{v}}}\right\rangle
	\nl&=\frac{1}{2}\sum_{w}\qty[V_{vw}-\sum_{\as}\tilde{\phi}_{\as vw}(\omega)]\left\langle \qty{\hat{P}_{u},\hat{Q}_{w}}\right\rangle
	\nl &=\sum_{w}\frac{1}{\pi}\qty[V_{vw}-\sum_{\as}\tilde{\phi}_{\as vw}(\omega)]\ii{0}{\infty}{\omega}\operatorname{Re}J^{QP}_{uw}(\omega)\coth\frac{\beta\omega}{2}
	\nl &=\ii{0}{\infty}{\omega}\mathbb{P}_{P_{u}Q_{v}}(\omega)\frac{\omega}{2}\coth\frac{\beta\omega}{2}
\end{align}
with
\begin{align}\label{C11}
	\mathbb{P}^{PQ}_{uv}(\omega)=\frac{2}{\pi\omega}\sum_{w}\qty[V_{vw}-\sum_{\as}\tilde{\phi}_{\as vw}(\omega)]\operatorname{Re}J^{PQ}_{uw}(\omega),
\end{align}
where $\bj^{PQ}(\omega)=\{J^{PQ}_{uv}(\omega)\}$ is the anti-Hermitian part of the matrix $\tilde{\cc}^{PQ}(\omega)=\{\tilde{\chi}^{PQ}_{uv}(\omega)=\ii{0}{\infty}{\omega}e^{i\omega t}\chi^{PQ}_{uv}(t)\}$. From \eq{chipq} we obtain $\bj^{PQ}(\omega)=-\omega\big[\bo^{-1}\tilde{\cc}^{QQ}(\omega)+\tilde{\cc}^{QQ\dagger}(\omega)\bo^{-1}\big]/2$, which also gives us 
\begin{align}\label{C12}
	\ii{0}{\infty}{\omega}\mathbb{P}_{P_{u}Q_{v}}(\omega)=0,
\end{align}
since $\ii{0}{\infty}{\omega}\tilde{\cc}^{QQ}(\omega)=\bm{0}$. 
Summarizing Eqs.\,(\ref{C1a}), (\ref{C3}), (\ref{C4}), (\ref{C6}), (\ref{C7}), (\ref{C9}), (\ref{C10}), (\ref{C12}) and the results of the cases of $\hat{X}_{i}=\hat{X}_{j}$ in the main text, we finally obtain \eqtwo{18}{19}. 

\section{Alternative approach for the energy}\label{appd}
To obtain \eqtwo{final}{B} for the multi-mode case, we set $N=n_{\s}+\sum_{\as}n_{\as}$ 
and follow the procedures in \citeref{Ford1988}:

(\uppercase\expandafter{\romannumeral 1}) From a normal-mode analysis 
we obtain the following relation bewteen $\tilde{\cc}^{QQ}(\omega)$ and all the normal frequencies $\{\overline{\omega}_{r}\}$:
\begin{align}\label{28}
	\det \tilde{\cc}^{QQ}(\omega)=(-1)^{n_{\s}}\det\bo\frac{\prod\limits_{\as j}(\omega^{2}-\omega^{2}_{\as j})}{\prod\limits_{r=1}^{N}(\omega^{2}-\overline{\omega}^{2}_{r})},
\end{align}
whose derivation is presented hereafter.

Applying a similar treatment as in \App{appa} to $\hQ_{u}(t)$ and $\hP_{u}(t)$, we obtain 
\begin{align}\label{ddotQ}
	\ddot{\hQ}_{u}(t)+\sum_{v}\Omega_{v}V_{uv}\hQ_{u}(t)+\sum_{\as j}\Omega_{u}c_{\as uj}\hx_{\as j}(t)=0.
\end{align}
Let 
\begin{align}\label{D1}
	\hQ_{u}(t)=\overline{Q}_{u}(\overline{\omega})e^{-i\overline{\omega} t},\quad
	\hx_{\as j}(t)=\overline{x}_{\as j}(\overline{\omega})e^{-i\overline{\omega} t},
\end{align}
where $\overline{\omega}$ is the normal frequency. To perform a normal-mode analysis, we put \eq{D1} into \eqtwo{ddotx}{ddotQ}, finding
\begin{subequations}
	\begin{align}
		&-\overline{\omega}^{2}\overline{Q}_{u}(\overline{\omega})+\sum_{v}\Omega_{v}V_{uv}\overline{Q}_{v}(\overline{\omega})+\sum_{\as j}\Omega_{u}c_{\as uj}\overline{x}_{\as j}(\overline{\omega})=0\label{D3a},\\
		&-\overline{\omega}^{2}\overline{x}_{\as j}(\overline{\omega})+\overline{\omega}^{2}_{\as j}\overline{x}_{\as j}(\overline{\omega})+\sum_{v}\frac{c_{\as uj}}{m_{\as j}}\overline{Q}_{u}(\overline{\omega})=0\label{D3b}.
	\end{align}
\end{subequations}
Solving \eq{D3b} for $\overline{x}_{\as j}(\overline{\omega})$, and substituting it into \eq{D3a}, we obtain
\begin{align}
	\sum_{v}\bigg[-\overline{\omega}^{2}\delta_{uv}+\Omega_{v}\delta_{uv}+\sum_{\as j}\Omega_{u}\frac{c_{\as uj}c_{\as vj}}{m_{\as j}(\overline{\omega}^{2}-\omega^{2}_{\as j})} \bigg]\overline{Q}_{v}(\overline{\omega})=0.
\end{align}
To obtain nontrivial normal frequencies, we require \cf{A5}
\begin{align}
	\det[\bo\bv-\overline{\omega}^{2}\bi-\bo\sum_{\as}\bp_{\as}(\overline{\omega})]=0,
\end{align}
which is also an equation for all the normal frequencies $\{\overline{\omega}\}$. As a function with respect to $\omega^{2}$ [cf.\,Eqs.\,(\ref{cc}) and (\ref{A5})], $\det\tilde{\cc}^{QQ}(\omega)$ has singular points at all $\{\omega^{2}_{r}\}$ while zero point at all $\{\omega^{2}_{\as j}\}$. Therefore we can write $\det\tilde{\cc}^{QQ}(\omega)$ in the form of \eq{28}. Note that $n_{\s}$ is the dimension of the matrix and the factor $(-1)^{n_{\s}}$ comes from the change of sign in the determinant.

\begin{figure}[h]
	\centering
	\begin{subfigure}[b]{0.5\textwidth}
		\includegraphics[width=\textwidth]{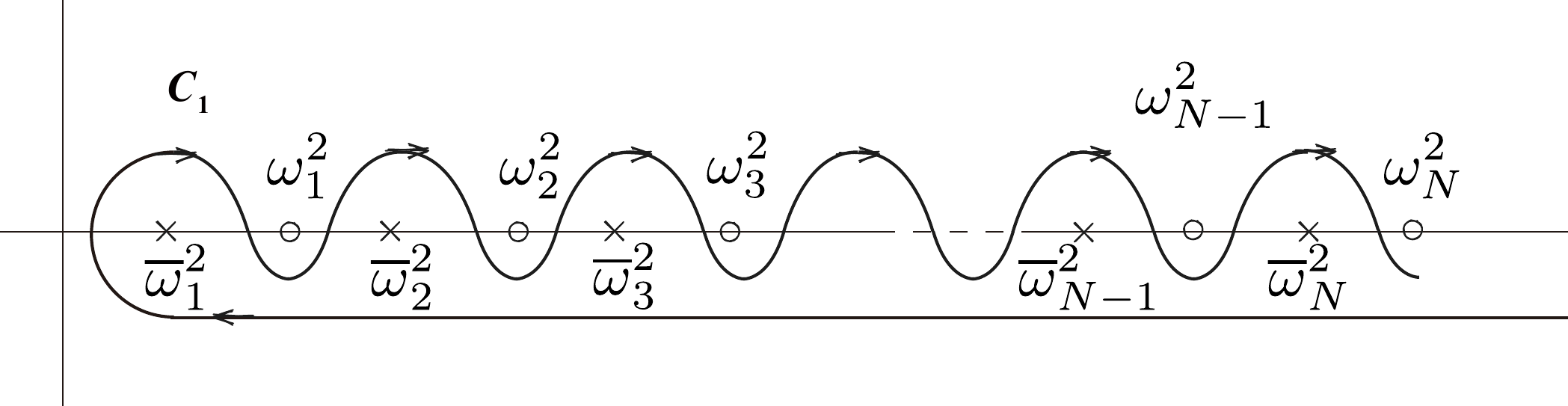}
		\caption{The contour $C_{1}$.}
		\label{C1}
	\end{subfigure}
	\begin{subfigure}[b]{0.5\textwidth}
		\includegraphics[width=\textwidth]{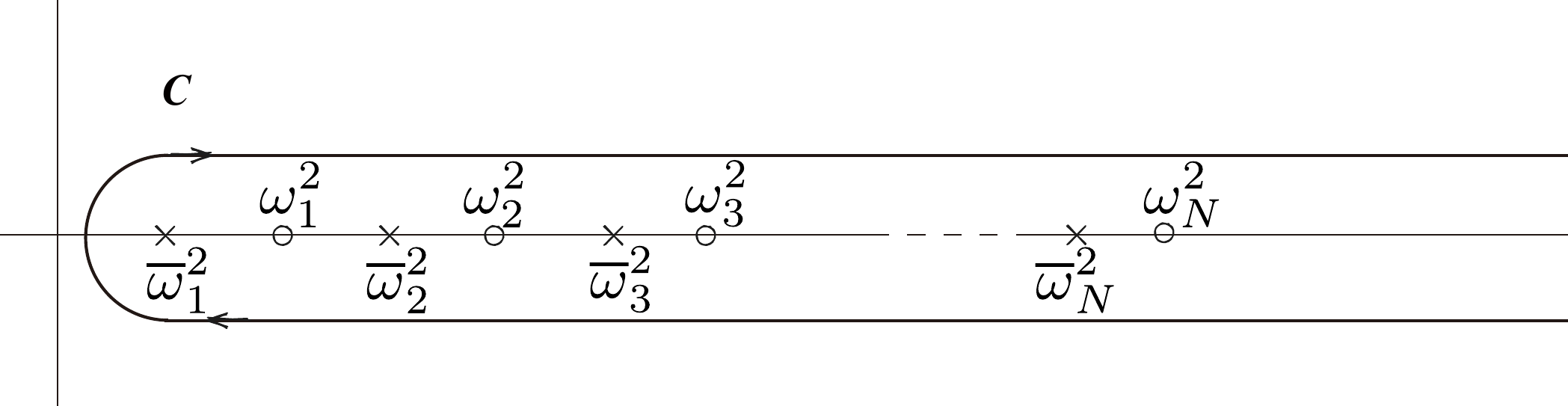}
		\caption{The contour $C$.}
		\label{C}
	\end{subfigure}
	\caption{Plots of contour $C_{1}$ and $C$. Here $\{\omega^{2}_{1},\omega^{2}_{2},...\}$ denote elements in $\{\omega^{2}_{\as j}\}$}
\end{figure}

(\uppercase\expandafter{\romannumeral 2}) Once we denote $\mathcal{A}(z)\equiv \det\tilde{\cc}^{QQ}(z^{1/2})$, mathematically it is easy to know 
\begin{align}\label{29}
	\der{z}[\cA(z)]^{-1}\bigg|_{z=\overline{\omega}^{2}_{r}}=\operatorname{Res}[\cA(z),\overline{\omega}^{2}_{r}]^{-1},
\end{align}
which helps us to recast \eq{UT} as
\begin{align}\label{30}
	U_{\T}(\beta)=\sum_{r=1}^{N}u(\overline{\omega}_{r},\beta)	\der{z}[\cA(z)]^{-1}\bigg|_{z=\overline{\omega}^{2}_{r}}\operatorname{Res}[\cA(z),\overline{\omega}^{2}_{r}].
\end{align}
By the residue theorem we further write \eq{30} as
\begin{align}
	U_{\T}(\beta)=-\frac{1}{2\pi i}\int_{C_{1}}\mathrm{d}z\,u(z^{1/2},\beta)\der{z}[\mathcal{A}(z)]^{-1}\cA(z),
\end{align}
where the contour $C_{1}$ is shown in \Fig{C1}.

(\uppercase\expandafter{\romannumeral 3}) In order to change the contour $C_{1}$ into $C$ (see \Fig{C}), we need to  consider a set of 
clockwise circle contours $C_{\as j}: |z-\omega^{2}_{\as j}|=\epsilon$. Since  $C=C_{1}+\sum_{\as j}C_{\as j}$ and we have
\begin{align}\label{32}
	&\sum_{\as j}\int_{|z-\omega^{2}_{\as j}|=\epsilon}\mathrm{d}z\,u(z^{1/2},\beta)\der{z}[\cA(z)]^{-1}\cA(z)
	\nl &\quad=2\pi i n_{\s}\sum_{\as j}\frac{\omega_{\as j}}{2}\coth\frac{\beta\omega_{\as j}}{2}.
\end{align}
Derivation of \eq{32} is straightforward as long as we remember \cf{cc}
\begin{align}\label{D6}
	\cA(z)&\equiv\det\tilde{\cc}^{QQ}(z^{1/2})\nl &=\det\bo/\det\big[\bo\bv-z\bi-\bo\sum_{\as}\bp_{\as}(z^{1/2})\big]\nl & \coloneq(-1)^{n_{\s}}\det\bo/\det\bC(z),
\end{align}
where we define
\begin{align}
	\bC(z)\equiv z\bi-\bo\bv-\bo\sum_{\as j}\bc_{\as j}\frac{\omega_{\as j}}{z-\omega^{2}_{\as j}}
\end{align}  for simplification.
Thus,
\begin{align}\label{S21}
	&\quad\der{z}[\cA(z)]^{-1}\cA(z)\nl &=(-1)^{n_{\s}}(\det\bo)^{-1}\der{z}\det\bC(z)\times (-1)^{n_{\s}}\det\bo/\det\bC(z)
	\nl	&=\det\bC(z)\operatorname{tr}\big[\bC^{-1}(z)\der{z}\bC(z)\big]/\det\bC(z)
	\nl 	&=\operatorname{tr}\big[\bC^{-1}(z)\der{z}\bC(z)\big].
\end{align}
By noting that
\begin{align}
	\der{z}\det\bC(z)=\det\bC(z)\operatorname{tr}\big[\bC^{-1}(z)\der{z}\bC(z)\big],
\end{align}
we find
\begin{align}\label{S23}
	&\quad\quad\bC^{-1}(z)\der{z}\bC(z)
	\nl &=\bigg[z\bi-\bo\bv-\bo\sum_{\as j}\bc_{\as j}\frac{\omega_{\as j}}{z-\omega^{2}_{\as j}}\bigg]^{-1}\bigg[\bi
	+\bo\nl &\quad\times\sum_{\as j}\bc_{\as j}\frac{\omega_{\as j}}{(z-\omega^{2}_{\as j})^{2}}\bigg]
	\nl &=\bigg[z\bi-\bo\bv-\bo\sum_{\as j}\bc_{\as j}\frac{\omega_{\as j}}{z-\omega^{2}_{\as j}}\bigg]^{-1}
	\times\frac{1}{(z-\omega^{2}_{\beta k})^{2}}\nl &\quad\times\bigg[(z-\omega^{2}_{\beta k})^{2}\bi+\bo\bc_{\beta k}\omega_{\beta
		k}+\bo\sum_{\as j\neq
		\beta k}\bc_{\as j}\omega_{\as j}\frac{(z-\omega^{2}_{\beta k})^{2}}{(z-\omega^{2}_{\as j})^{2}}\bigg]
	\nl &=\bigg[z(z-\omega^{2}_{\beta k})^{2}\bi-\bo\bv(z-\omega^{2}_{\beta k})^{2}-\bo\bc_{\beta k}\omega_{\beta k}(z-\omega^{2}_{\beta k})
	\nl &\quad-\bo\sum_{\as j\neq \beta k}\bc_{\as j}\omega_{\as j}\frac{(z-\omega^{2}_{\beta k})^{2}}{z-\omega^{2}_{\as j}}\bigg]^{-1}
	\nl &\quad\times\bigg[(z-\omega^{2}_{\beta k})^{2}\bi+\bo\bc_{\beta k}\omega_{\beta
		k}+\bo\sum_{\as j\neq
		\beta k}\bc_{\as j}\omega_{\as j}\frac{(z-\omega^{2}_{\beta k})^{2}}{(z-\omega^{2}_{\as j})^{2}}\bigg]
	\nl &\overset{\lim\limits_{z\rightarrow \omega^{2}_{\beta k}}}{=}\big[-\bo\bc_{\beta k}\omega_{\beta k}(z-\omega^{2}_{\beta k})\big]^{-1}\big(\bo\bc_{\beta k}\omega_{\beta k}\big)
	\nl &=-\frac{1}{z-\omega^{2}_{\beta k}}\bi.
\end{align}
Evaluating \eq{32} using \eqs{S21} and (\ref{S23}) gives us \eq{32}.
By \eq{UB} we deduce the following expression:
\begin{align}\label{UTz}
	U_{\T}(\beta)-n_{\s}U_{\B}(\beta)=-\frac{1}{2\pi i}\int_{C}\mathrm{d}z\,u(z^{1/2},\beta)\der{z}[\mathcal{A}(z)]^{-1}\cA(z).
\end{align}

(\uppercase\expandafter{\romannumeral 4}) First we set $z=\omega^{2}$ in \eq{UTz}. Then by noticing that $\tilde{\cc}^{QQ}(\omega)=[\tilde{\cc}^{QQ}(-\omega)]^{*}$ \cite{Yan2005} and the function $u(\omega,\beta)$ is even with repect to $\omega$, we simplify \eq{UTz} to obtain the final remarkable expression:
\begin{align}\label{D17}
	U_{\T}(\beta)-n_{\s}U_{\B}(\beta)=\frac{1}{\pi}\ii{0}{\infty}{\omega}u(\omega,\beta)\operatorname{Im}\der{\omega}\ln\det\cc^{QQ}(\omega).
\end{align}
\Eq{D17} can also be recast to the form of equipartition theorem:
\begin{align}
	U_{\T}(\beta)-n_{\s}U_{\B}(\beta)=\ii{0}{\infty}{\omega}\mathbb{B}(\omega)\frac{\omega}{2}\coth\frac{\beta\omega}{2},
\end{align}
where
\begin{align}
	\mathbb{B}(\omega)=\frac{1}{\pi n_{\s}}\operatorname{Im}\der{\omega}\ln\det\tilde{\cc}^{QQ}(\omega).
\end{align}
Therefore we obtain \eqtwo{final}{B}, which 
 reduce to \eq{Us} when $n_{\s}=1$.

\section{Proof of \eq{301} in the main text}\label{appe}
In this section we give a proof of \eq{301}.
From Eqs.\,\,(\ref{PQ}),\,\,(\ref{PP}) and (\ref{P}) we obtain
\begin{align}\label{S49}
	\mathbb{P}(\omega)&=\frac{1}{2n_{\s}}\sum_{u}\big[ \mathbb{P}_{Q_uQ_u}(\omega)+\mathbb{P}_{P_uP_u}(\omega)\big]
	\nl &
	=\frac{1}{2n_{\s}}\sum_{u}\bigg\{ \frac{2}{\pi\omega}\sum_{v}\bigg[V_{uv}-\sum_{\as}\tilde{\phi}_{\as uv}(0)\bigg]J^{QQ}_{uv}(\omega)
	\nl &\quad+\frac{2\omega}{\pi\Omega_{u}}J^{QQ}_{uu}(\omega)\bigg\}
	\nl &
	=\frac{1}{\pi n_{\s}\omega} \operatorname{tr}\bigg\{\bigg[\bv-\sum_{\as}\bp_{\as}(0)+\omega^{2}\bo^{-1}\bigg]\bj^{QQ}(\omega)\bigg\}
	\nl &
	=\frac{1}{\pi n_{\s}\omega} \operatorname{tr}\bigg\{\bigg[\bv-\sum_{\as}\bp_{\as}(0)+\omega^{2}\bo^{-1}\bigg]
	\nl &\quad\times\operatorname{Im}\bigg
	[\bo\bv-\omega^2 \bi-\bo \sum_\alpha \bp_\alpha(\omega)\bigg]^{-1}\bigg\}.
\end{align}
In the weak-coupling limit, all $c_{\as u j }\rightarrow 0$, and therefore $\bm{\phi}_{\as}(\omega)\rightarrow \bm{0}$ \cf{A5}. Equation\,\,(\ref{S49}) becomes
\begin{align}\label{E2}
	\mathbb{P}(\omega)&=	\frac{1}{\pi n_{\s}\omega} \operatorname{tr}\big[\big(\bv+\omega^{2}\bo^{-1}\big)\operatorname{Im}\big
	(\bo\bv-\omega^2 \bi\big)^{-1}\big]
	\nl &=
	\begin{cases}
		0, & \text{if $\bo\bv-\omega^{2}\bi$ is invertible},  \\
		\infty, & \text{if $\bo\bv-\omega^{2}\bi$ is not invertible}, \\
	\end{cases}
\end{align}
but we still know from the normalization of $\mathbb{P}_{Q_uQ_u}(\omega)$ and $\mathbb{P}_{P_uP_u}(\omega)$ that 
\begin{align}\label{E3}
	\ii{0}{\infty}{\omega}\mathbb{P}(\omega)=1.
\end{align}
\Eq{E2} is always zero except for the squre root of the eigenvalues of the matrix $\bo\bv$. Considering \eq{E3} and the definition for Dirac delta function, we obtain
\begin{align}
	\lim\limits_{\{c_{\as uj}\}\rightarrow \{0\}}\mathbb{P}(\omega)=\frac{1}{n_{\s}}\sum_{\omega^{2}_{i}\in \operatorname{spec}(\bo\bv)}\delta(\omega-\omega_{i}) \quad \omega>0,
\end{align}
where the summation is over all the square root of eigenvalues of the matrix $\bo\bv$  (considering multiplicity). The coefficients $1/n_{\s}$ comes from the numbers of the positive eigenvalues of $\bo\bv$, which equals $n_{\s}$
due to the fact that $\bo\bv$ and $\sqrt{\bo}\bv\sqrt{\bo}$ are similar and the latter is a positive definite matrix of dimension $n_{\s}$.
Since $\bj^{QQ}(\omega)$ is odd \cite{Yan2005}, it follows that $\mathbb{P}(\omega)$ is an even function, as indicated by the penultimate equality in \eq{S49}. With this we obtain
the \eq{301}.


\begin{comment}
\bibliography{ref.bib}	

\begin{thebibliography}{31}%
	\makeatletter
	\providecommand \@ifxundefined [1]{%
		\@ifx{#1\undefined}
	}%
	\providecommand \@ifnum [1]{%
		\ifnum #1\expandafter \@firstoftwo
		\else \expandafter \@secondoftwo
		\fi
	}%
	\providecommand \@ifx [1]{%
		\ifx #1\expandafter \@firstoftwo
		\else \expandafter \@secondoftwo
		\fi
	}%
	\providecommand \natexlab [1]{#1}%
	\providecommand \enquote  [1]{``#1''}%
	\providecommand \bibnamefont  [1]{#1}%
	\providecommand \bibfnamefont [1]{#1}%
	\providecommand \citenamefont [1]{#1}%
	\providecommand \href@noop [0]{\@secondoftwo}%
	\providecommand \href [0]{\begingroup \@sanitize@url \@href}%
	\providecommand \@href[1]{\@@startlink{#1}\@@href}%
	\providecommand \@@href[1]{\endgroup#1\@@endlink}%
	\providecommand \@sanitize@url [0]{\catcode `\\12\catcode `\$12\catcode
		`\&12\catcode `\#12\catcode `\^12\catcode `\_12\catcode `\%12\relax}%
	\providecommand \@@startlink[1]{}%
	\providecommand \@@endlink[0]{}%
	\providecommand \url  [0]{\begingroup\@sanitize@url \@url }%
	\providecommand \@url [1]{\endgroup\@href {#1}{\urlprefix }}%
	\providecommand \urlprefix  [0]{URL }%
	\providecommand \Eprint [0]{\href }%
	\providecommand \doibase [0]{https://doi.org/}%
	\providecommand \selectlanguage [0]{\@gobble}%
	\providecommand \bibinfo  [0]{\@secondoftwo}%
	\providecommand \bibfield  [0]{\@secondoftwo}%
	\providecommand \translation [1]{[#1]}%
	\providecommand \BibitemOpen [0]{}%
	\providecommand \bibitemStop [0]{}%
	\providecommand \bibitemNoStop [0]{.\EOS\space}%
	\providecommand \EOS [0]{\spacefactor3000\relax}%
	\providecommand \BibitemShut  [1]{\csname bibitem#1\endcsname}%
	\let\auto@bib@innerbib\@empty
	%</preamble>
	\bibitem [{\citenamefont {Dattagupta}\ \emph {et~al.}(2010)\citenamefont
		{Dattagupta}, \citenamefont {Kumar}, \citenamefont {Sinha},\ and\
		\citenamefont {Sreeram}}]{Dattagupta2010}%
	\BibitemOpen
	\bibfield  {author} {\bibinfo {author} {\bibfnamefont {S.}~\bibnamefont
			{Dattagupta}}, \bibinfo {author} {\bibfnamefont {J.}~\bibnamefont {Kumar}},
		\bibinfo {author} {\bibfnamefont {S.}~\bibnamefont {Sinha}},\ and\ \bibinfo
		{author} {\bibfnamefont {P.~A.}\ \bibnamefont {Sreeram}},\ }\bibfield
	{title} {\bibinfo {title} {Dissipative quantum systems and the heat
			capacity},\ }\href {https://doi.org/10.1103/physreve.81.031136} {\bibfield
		{journal} {\bibinfo  {journal} {Physical Review E}\ }\textbf {\bibinfo
			{volume} {81}},\ \bibinfo {pages} {031136} (\bibinfo {year}
		{2010})}\BibitemShut {NoStop}%
	\bibitem [{\citenamefont {Campisi}(2007)}]{Campisi2007}%
	\BibitemOpen
	\bibfield  {author} {\bibinfo {author} {\bibfnamefont {M.}~\bibnamefont
			{Campisi}},\ }\bibfield  {title} {\bibinfo {title} {On the limiting cases of
			nonextensive thermostatistics},\ }\href
	{https://doi.org/10.1016/j.physleta.2007.01.082} {\bibfield  {journal}
		{\bibinfo  {journal} {Physics Letters A}\ }\textbf {\bibinfo {volume}
			{366}},\ \bibinfo {pages} {335} (\bibinfo {year} {2007})}\BibitemShut
	{NoStop}%
	\bibitem [{\citenamefont {Koide}\ and\ \citenamefont
		{Kodama}(2011)}]{Koide2011}%
	\BibitemOpen
	\bibfield  {author} {\bibinfo {author} {\bibfnamefont {T.}~\bibnamefont
			{Koide}}\ and\ \bibinfo {author} {\bibfnamefont {T.}~\bibnamefont {Kodama}},\
	}\bibfield  {title} {\bibinfo {title} {{Thermodynamic laws and equipartition
				theorem in relativistic Brownian motion}},\ }\href
	{https://doi.org/10.1103/physreve.83.061111} {\bibfield  {journal} {\bibinfo
			{journal} {Physical Review E}\ }\textbf {\bibinfo {volume} {83}},\ \bibinfo
		{pages} {061111} (\bibinfo {year} {2011})}\BibitemShut {NoStop}%
	\bibitem [{\citenamefont {Abreu}\ \emph
		{et~al.}(2020{\natexlab{a}})\citenamefont {Abreu}, \citenamefont {Neto},
		\citenamefont {Barboza}, \citenamefont {Mendes},\ and\ \citenamefont
		{Soares}}]{Abreu2020}%
	\BibitemOpen
	\bibfield  {author} {\bibinfo {author} {\bibfnamefont {E.~M.~C.}\
			\bibnamefont {Abreu}}, \bibinfo {author} {\bibfnamefont {J.~A.}\ \bibnamefont
			{Neto}}, \bibinfo {author} {\bibfnamefont {E.~M.}\ \bibnamefont {Barboza}},
		\bibinfo {author} {\bibfnamefont {A.~C.~R.}\ \bibnamefont {Mendes}},\ and\
		\bibinfo {author} {\bibfnamefont {B.~B.}\ \bibnamefont {Soares}},\ }\bibfield
	{title} {\bibinfo {title} {{On the equipartition theorem and black holes
				non-Gaussian entropies}},\ }\href {https://doi.org/10.1142/s0217732320502661}
	{\bibfield  {journal} {\bibinfo  {journal} {Modern Physics Letters A}\
		}\textbf {\bibinfo {volume} {35}},\ \bibinfo {pages} {2050266} (\bibinfo
		{year} {2020}{\natexlab{a}})}\BibitemShut {NoStop}%
	\bibitem [{\citenamefont {Abreu}\ \emph
		{et~al.}(2020{\natexlab{b}})\citenamefont {Abreu}, \citenamefont {Neto},\
		and\ \citenamefont {Barboza}}]{Abreu2020a}%
	\BibitemOpen
	\bibfield  {author} {\bibinfo {author} {\bibfnamefont {E.~M.~C.}\
			\bibnamefont {Abreu}}, \bibinfo {author} {\bibfnamefont {J.~A.}\ \bibnamefont
			{Neto}},\ and\ \bibinfo {author} {\bibfnamefont {E.~M.}\ \bibnamefont
			{Barboza}},\ }\bibfield  {title} {\bibinfo {title} {Barrow{\textquotesingle}s
			black hole entropy and the equipartition theorem},\ }\href
	{https://doi.org/10.1209/0295-5075/130/40005} {\bibfield  {journal} {\bibinfo
			{journal} {{EPL} (Europhysics Letters)}\ }\textbf {\bibinfo {volume}
			{130}},\ \bibinfo {pages} {40005} (\bibinfo {year}
		{2020}{\natexlab{b}})}\BibitemShut {NoStop}%
	\bibitem [{\citenamefont {Barboza}\ \emph {et~al.}(2015)\citenamefont
		{Barboza}, \citenamefont {da~C.~Nunes}, \citenamefont {Abreu},\ and\
		\citenamefont {Neto}}]{Barboza2015}%
	\BibitemOpen
	\bibfield  {author} {\bibinfo {author} {\bibfnamefont {E.~M.}\ \bibnamefont
			{Barboza}}, \bibinfo {author} {\bibfnamefont {R.}~\bibnamefont
			{da~C.~Nunes}}, \bibinfo {author} {\bibfnamefont {E.~M.}\ \bibnamefont
			{Abreu}},\ and\ \bibinfo {author} {\bibfnamefont {J.~A.}\ \bibnamefont
			{Neto}},\ }\bibfield  {title} {\bibinfo {title} {{Dark energy models through
				nonextensive Tsallis' statistics}},\ }\href
	{https://doi.org/10.1016/j.physa.2015.05.002} {\bibfield  {journal} {\bibinfo
			{journal} {Physica A: Statistical Mechanics and its Applications}\ }\textbf
		{\bibinfo {volume} {436}},\ \bibinfo {pages} {301} (\bibinfo {year}
		{2015})}\BibitemShut {NoStop}%
	\bibitem [{\citenamefont {van~der Ziel}(1973)}]{Ziel1973}%
	\BibitemOpen
	\bibfield  {author} {\bibinfo {author} {\bibfnamefont {A.}~\bibnamefont
			{van~der Ziel}},\ }\bibfield  {title} {\bibinfo {title} {Equivalent circuit
			and equipartition theorem in ideal dielectric and ferroelectric capacitors},\
	}\href {https://doi.org/10.1063/1.1662366} {\bibfield  {journal} {\bibinfo
			{journal} {Journal of Applied Physics}\ }\textbf {\bibinfo {volume} {44}},\
		\bibinfo {pages} {1400} (\bibinfo {year} {1973})}\BibitemShut {NoStop}%
	\bibitem [{\citenamefont {Sarpeshkar}\ \emph {et~al.}(1993)\citenamefont
		{Sarpeshkar}, \citenamefont {Delbruck},\ and\ \citenamefont
		{Mead}}]{Sarpeshkar1993}%
	\BibitemOpen
	\bibfield  {author} {\bibinfo {author} {\bibfnamefont {R.}~\bibnamefont
			{Sarpeshkar}}, \bibinfo {author} {\bibfnamefont {T.}~\bibnamefont
			{Delbruck}},\ and\ \bibinfo {author} {\bibfnamefont {C.}~\bibnamefont
			{Mead}},\ }\bibfield  {title} {\bibinfo {title} {White noise in {MOS}
			transistors and resistors},\ }\href {https://doi.org/10.1109/101.261888}
	{\bibfield  {journal} {\bibinfo  {journal} {{IEEE} Circuits and Devices
				Magazine}\ }\textbf {\bibinfo {volume} {9}},\ \bibinfo {pages} {23} (\bibinfo
		{year} {1993})}\BibitemShut {NoStop}%
	\bibitem [{\citenamefont {Matheny}\ \emph {et~al.}(2013)\citenamefont
		{Matheny}, \citenamefont {Villanueva}, \citenamefont {Karabalin},
		\citenamefont {Sader},\ and\ \citenamefont {Roukes}}]{Matheny2013}%
	\BibitemOpen
	\bibfield  {author} {\bibinfo {author} {\bibfnamefont {M.~H.}\ \bibnamefont
			{Matheny}}, \bibinfo {author} {\bibfnamefont {L.~G.}\ \bibnamefont
			{Villanueva}}, \bibinfo {author} {\bibfnamefont {R.~B.}\ \bibnamefont
			{Karabalin}}, \bibinfo {author} {\bibfnamefont {J.~E.}\ \bibnamefont
			{Sader}},\ and\ \bibinfo {author} {\bibfnamefont {M.~L.}\ \bibnamefont
			{Roukes}},\ }\bibfield  {title} {\bibinfo {title} {Nonlinear mode-coupling in
			nanomechanical systems},\ }\href {https://doi.org/10.1021/nl400070e}
	{\bibfield  {journal} {\bibinfo  {journal} {Nano Letters}\ }\textbf {\bibinfo
			{volume} {13}},\ \bibinfo {pages} {1622} (\bibinfo {year}
		{2013})}\BibitemShut {NoStop}%
	\bibitem [{\citenamefont {Bialas}\ \emph {et~al.}(2019)\citenamefont {Bialas},
		\citenamefont {Spiechowicz},\ and\ \citenamefont {{\L}uczka}}]{Bialas2019}%
	\BibitemOpen
	\bibfield  {author} {\bibinfo {author} {\bibfnamefont {P.}~\bibnamefont
			{Bialas}}, \bibinfo {author} {\bibfnamefont {J.}~\bibnamefont
			{Spiechowicz}},\ and\ \bibinfo {author} {\bibfnamefont {J.}~\bibnamefont
			{{\L}uczka}},\ }\bibfield  {title} {\bibinfo {title} {Quantum analogue of
			energy equipartition theorem},\ }\href
	{https://doi.org/10.1088/1751-8121/ab03f2} {\bibfield  {journal} {\bibinfo
			{journal} {Journal of Physics A: Mathematical and Theoretical}\ }\textbf
		{\bibinfo {volume} {52}},\ \bibinfo {pages} {15LT01} (\bibinfo {year}
		{2019})}\BibitemShut {NoStop}%
	\bibitem [{\citenamefont {Ghosh}(2023)}]{Ghosh_2023}%
	\BibitemOpen
	\bibfield  {author} {\bibinfo {author} {\bibfnamefont {A.}~\bibnamefont
			{Ghosh}},\ }\bibfield  {title} {\bibinfo {title} {Generalised energy
			equipartition in electrical circuits},\ }\bibfield  {journal} {\bibinfo
		{journal} {Pramana}\ }\textbf {\bibinfo {volume} {97}},\ \href
	{https://doi.org/10.1007/s12043-023-02553-w} {10.1007/s12043-023-02553-w}
	(\bibinfo {year} {2023})\BibitemShut {NoStop}%
	\bibitem [{\citenamefont {Kaur}\ \emph {et~al.}(2021)\citenamefont {Kaur},
		\citenamefont {Ghosh},\ and\ \citenamefont {Bandyopadhyay}}]{Kaur2021}%
	\BibitemOpen
	\bibfield  {author} {\bibinfo {author} {\bibfnamefont {J.}~\bibnamefont
			{Kaur}}, \bibinfo {author} {\bibfnamefont {A.}~\bibnamefont {Ghosh}},\ and\
		\bibinfo {author} {\bibfnamefont {M.}~\bibnamefont {Bandyopadhyay}},\
	}\bibfield  {title} {\bibinfo {title} {Quantum counterpart of energy
			equipartition theorem for a dissipative charged magneto-oscillator: Effect of
			dissipation, memory, and magnetic field},\ }\href
	{https://doi.org/10.1103/physreve.104.064112} {\bibfield  {journal} {\bibinfo
			{journal} {Physical Review E}\ }\textbf {\bibinfo {volume} {104}},\ \bibinfo
		{pages} {064112} (\bibinfo {year} {2021})}\BibitemShut {NoStop}%
	\bibitem [{\citenamefont {Kaur}\ \emph {et~al.}(2022)\citenamefont {Kaur},
		\citenamefont {Ghosh},\ and\ \citenamefont {Bandyopadhyay}}]{Kaur2022a}%
	\BibitemOpen
	\bibfield  {author} {\bibinfo {author} {\bibfnamefont {J.}~\bibnamefont
			{Kaur}}, \bibinfo {author} {\bibfnamefont {A.}~\bibnamefont {Ghosh}},\ and\
		\bibinfo {author} {\bibfnamefont {M.}~\bibnamefont {Bandyopadhyay}},\
	}\bibfield  {title} {\bibinfo {title} {{Partition of free energy for a
				Brownian quantum oscillator: Effect of dissipation and magnetic field}},\
	}\href {https://doi.org/10.1016/j.physa.2022.127466} {\bibfield  {journal}
		{\bibinfo  {journal} {Physica A: Statistical Mechanics and its Applications}\
		}\textbf {\bibinfo {volume} {599}},\ \bibinfo {pages} {127466} (\bibinfo
		{year} {2022})}\BibitemShut {NoStop}%
	\bibitem [{\citenamefont {Bialas}\ \emph {et~al.}(2018)\citenamefont {Bialas},
		\citenamefont {Spiechowicz},\ and\ \citenamefont {{\L}uczka}}]{Bialas2018}%
	\BibitemOpen
	\bibfield  {author} {\bibinfo {author} {\bibfnamefont {P.}~\bibnamefont
			{Bialas}}, \bibinfo {author} {\bibfnamefont {J.}~\bibnamefont
			{Spiechowicz}},\ and\ \bibinfo {author} {\bibfnamefont {J.}~\bibnamefont
			{{\L}uczka}},\ }\bibfield  {title} {\bibinfo {title} {Partition of energy for
			a dissipative quantum oscillator},\ }\bibfield  {journal} {\bibinfo
		{journal} {Scientific Reports}\ }\textbf {\bibinfo {volume} {8}},\ \href
	{https://doi.org/10.1038/s41598-018-34385-9} {10.1038/s41598-018-34385-9}
	(\bibinfo {year} {2018})\BibitemShut {NoStop}%
	\bibitem [{\citenamefont {Spiechowicz}\ \emph {et~al.}(2018)\citenamefont
		{Spiechowicz}, \citenamefont {Bialas},\ and\ \citenamefont
		{{\L}uczka}}]{Spiechowicz2018}%
	\BibitemOpen
	\bibfield  {author} {\bibinfo {author} {\bibfnamefont {J.}~\bibnamefont
			{Spiechowicz}}, \bibinfo {author} {\bibfnamefont {P.}~\bibnamefont
			{Bialas}},\ and\ \bibinfo {author} {\bibfnamefont {J.}~\bibnamefont
			{{\L}uczka}},\ }\bibfield  {title} {\bibinfo {title} {{Quantum partition of
				energy for a free Brownian particle: Impact of dissipation}},\ }\href
	{https://doi.org/10.1103/physreva.98.052107} {\bibfield  {journal} {\bibinfo
			{journal} {Physical Review A}\ }\textbf {\bibinfo {volume} {98}},\ \bibinfo
		{pages} {052107} (\bibinfo {year} {2018})}\BibitemShut {NoStop}%
	\bibitem [{\citenamefont {Spiechowicz}\ and\ \citenamefont
		{{\L}uczka}(2019)}]{Spiechowicz2019}%
	\BibitemOpen
	\bibfield  {author} {\bibinfo {author} {\bibfnamefont {J.}~\bibnamefont
			{Spiechowicz}}\ and\ \bibinfo {author} {\bibfnamefont {J.}~\bibnamefont
			{{\L}uczka}},\ }\bibfield  {title} {\bibinfo {title} {{On superstatistics of
				energy for a free quantum Brownian particle}},\ }\href
	{https://doi.org/10.1088/1742-5468/ab1c4e} {\bibfield  {journal} {\bibinfo
			{journal} {Journal of Statistical Mechanics: Theory and Experiment}\ }\textbf
		{\bibinfo {volume} {2019}},\ \bibinfo {pages} {064002} (\bibinfo {year}
		{2019})}\BibitemShut {NoStop}%
	\bibitem [{\citenamefont {Spiechowicz}\ and\ \citenamefont
		{{\L}uczka}(2021)}]{Spiechowicz2021}%
	\BibitemOpen
	\bibfield  {author} {\bibinfo {author} {\bibfnamefont {J.}~\bibnamefont
			{Spiechowicz}}\ and\ \bibinfo {author} {\bibfnamefont {J.}~\bibnamefont
			{{\L}uczka}},\ }\bibfield  {title} {\bibinfo {title} {{Energy of a free
				Brownian particle coupled to thermal vacuum}},\ }\bibfield  {journal}
	{\bibinfo  {journal} {Scientific Reports}\ }\textbf {\bibinfo {volume}
		{11}},\ \href {https://doi.org/10.1038/s41598-021-83617-y}
	{10.1038/s41598-021-83617-y} (\bibinfo {year} {2021})\BibitemShut {NoStop}%
	\bibitem [{\citenamefont {Kaur}\ \emph {et~al.}(2023)\citenamefont {Kaur},
		\citenamefont {Ghosh},\ and\ \citenamefont {Bandyopadhyay}}]{Kaur2023}%
	\BibitemOpen
	\bibfield  {author} {\bibinfo {author} {\bibfnamefont {J.}~\bibnamefont
			{Kaur}}, \bibinfo {author} {\bibfnamefont {A.}~\bibnamefont {Ghosh}},\ and\
		\bibinfo {author} {\bibfnamefont {M.}~\bibnamefont {Bandyopadhyay}},\
	}\bibfield  {title} {\bibinfo {title} {Partition of kinetic energy and
			magnetic moment in dissipative diamagnetism},\ }\href
	{https://doi.org/10.1016/j.physa.2023.128993} {\bibfield  {journal} {\bibinfo
			{journal} {Physica A: Statistical Mechanics and its Applications}\ }\textbf
		{\bibinfo {volume} {625}},\ \bibinfo {pages} {128993} (\bibinfo {year}
		{2023})}\BibitemShut {NoStop}%
	\bibitem [{\citenamefont {{\L}uczka}(2020)}]{luczka2020quantum}%
	\BibitemOpen
	\bibfield  {author} {\bibinfo {author} {\bibfnamefont {J.}~\bibnamefont
			{{\L}uczka}},\ }\bibfield  {title} {\bibinfo {title} {Quantum counterpart of
			classical equipartition of energy},\ }\href
	{https://doi.org/10.1007/s10955-020-02557-5} {\bibfield  {journal} {\bibinfo
			{journal} {Journal of Statistical Physics}\ }\textbf {\bibinfo {volume}
			{179}},\ \bibinfo {pages} {839} (\bibinfo {year} {2020})}\BibitemShut
	{NoStop}%
	\bibitem [{\citenamefont {Magnano}\ and\ \citenamefont
		{Valsesia}(2020)}]{Magnano2020}%
	\BibitemOpen
	\bibfield  {author} {\bibinfo {author} {\bibfnamefont {G.}~\bibnamefont
			{Magnano}}\ and\ \bibinfo {author} {\bibfnamefont {B.}~\bibnamefont
			{Valsesia}},\ }\bibfield  {title} {\bibinfo {title} {On the generalised
			equipartition law},\ }\bibfield  {journal} {\bibinfo  {journal} {Annals of
			Physics 427 (2021), 168416}\ }\href
	{https://doi.org/10.1016/j.aop.2021.168416} {10.1016/j.aop.2021.168416}
	(\bibinfo {year} {2020}),\ \Eprint {https://arxiv.org/abs/2009.02518}
	{arXiv:2009.02518 [math-ph]} \BibitemShut {NoStop}%
	\bibitem [{\citenamefont {Caldeira}\ and\ \citenamefont
		{Leggett}(1981)}]{Caldeira1981}%
	\BibitemOpen
	\bibfield  {author} {\bibinfo {author} {\bibfnamefont {A.~O.}\ \bibnamefont
			{Caldeira}}\ and\ \bibinfo {author} {\bibfnamefont {A.~J.}\ \bibnamefont
			{Leggett}},\ }\bibfield  {title} {\bibinfo {title} {Influence of dissipation
			on quantum tunneling in macroscopic systems},\ }\href
	{https://doi.org/10.1103/physrevlett.46.211} {\bibfield  {journal} {\bibinfo
			{journal} {Physical Review Letters}\ }\textbf {\bibinfo {volume} {46}},\
		\bibinfo {pages} {211} (\bibinfo {year} {1981})}\BibitemShut {NoStop}%
	\bibitem [{\citenamefont {Tong}\ \emph {et~al.}(2023)\citenamefont {Tong},
		\citenamefont {Gong}, \citenamefont {Wang}, \citenamefont {Xu},\ and\
		\citenamefont {Yan}}]{Tong2023}%
	\BibitemOpen
	\bibfield  {author} {\bibinfo {author} {\bibfnamefont {X.-H.}\ \bibnamefont
			{Tong}}, \bibinfo {author} {\bibfnamefont {H.}~\bibnamefont {Gong}}, \bibinfo
		{author} {\bibfnamefont {Y.}~\bibnamefont {Wang}}, \bibinfo {author}
		{\bibfnamefont {R.-X.}\ \bibnamefont {Xu}},\ and\ \bibinfo {author}
		{\bibfnamefont {Y.}~\bibnamefont {Yan}},\ }\bibfield  {title} {\bibinfo
		{title} {{Multimode Brownian oscillators: Exact solutions to heat
				transport}},\ }\bibfield  {journal} {\bibinfo  {journal} {The Journal of
			Chemical Physics}\ }\textbf {\bibinfo {volume} {159}},\ \href
	{https://doi.org/10.1063/5.0157186} {10.1063/5.0157186} (\bibinfo {year}
	{2023})\BibitemShut {NoStop}%
	\bibitem [{\citenamefont {Ford}\ \emph {et~al.}(1988)\citenamefont {Ford},
		\citenamefont {Lewis},\ and\ \citenamefont
		{O{\textquotesingle}Connell}}]{Ford1988}%
	\BibitemOpen
	\bibfield  {author} {\bibinfo {author} {\bibfnamefont {G.~W.}\ \bibnamefont
			{Ford}}, \bibinfo {author} {\bibfnamefont {J.~T.}\ \bibnamefont {Lewis}},\
		and\ \bibinfo {author} {\bibfnamefont {R.~F.}\ \bibnamefont
			{O{\textquotesingle}Connell}},\ }\bibfield  {title} {\bibinfo {title}
		{{Independent oscillator model of a heat bath: Exact diagonalization of the
				Hamiltonian}},\ }\href {https://doi.org/10.1007/bf01011565} {\bibfield
		{journal} {\bibinfo  {journal} {Journal of Statistical Physics}\ }\textbf
		{\bibinfo {volume} {53}},\ \bibinfo {pages} {439} (\bibinfo {year}
		{1988})}\BibitemShut {NoStop}%
	\bibitem [{\citenamefont {Ghosh}\ \emph {et~al.}(2023)\citenamefont {Ghosh},
		\citenamefont {Bandyopadhyay}, \citenamefont {Dattagupta},\ and\
		\citenamefont {Gupta}}]{Ghosh2023}%
	\BibitemOpen
	\bibfield  {author} {\bibinfo {author} {\bibfnamefont {A.}~\bibnamefont
			{Ghosh}}, \bibinfo {author} {\bibfnamefont {M.}~\bibnamefont
			{Bandyopadhyay}}, \bibinfo {author} {\bibfnamefont {S.}~\bibnamefont
			{Dattagupta}},\ and\ \bibinfo {author} {\bibfnamefont {S.}~\bibnamefont
			{Gupta}},\ }\bibfield  {title} {\bibinfo {title} {{Quantum Brownian Motion: A
				Review}},\ }\href@noop {} {\  (\bibinfo {year} {2023})},\ \Eprint
	{https://arxiv.org/abs/2306.02665} {arXiv:2306.02665 [cond-mat.stat-mech]}
	\BibitemShut {NoStop}%
	\bibitem [{\citenamefont {Callen}\ and\ \citenamefont
		{Welton}(1951)}]{Callen1951}%
	\BibitemOpen
	\bibfield  {author} {\bibinfo {author} {\bibfnamefont {H.~B.}\ \bibnamefont
			{Callen}}\ and\ \bibinfo {author} {\bibfnamefont {T.~A.}\ \bibnamefont
			{Welton}},\ }\bibfield  {title} {\bibinfo {title} {Irreversibility and
			generalized noise},\ }\href {https://doi.org/10.1103/physrev.83.34}
	{\bibfield  {journal} {\bibinfo  {journal} {Physical Review}\ }\textbf
		{\bibinfo {volume} {83}},\ \bibinfo {pages} {34} (\bibinfo {year}
		{1951})}\BibitemShut {NoStop}%
	\bibitem [{\citenamefont {Feynman}\ and\ \citenamefont
		{Vernon}(2000)}]{Feynman2000}%
	\BibitemOpen
	\bibfield  {author} {\bibinfo {author} {\bibfnamefont {R.}~\bibnamefont
			{Feynman}}\ and\ \bibinfo {author} {\bibfnamefont {F.}~\bibnamefont
			{Vernon}},\ }\bibfield  {title} {\bibinfo {title} {The theory of a general
			quantum system interacting with a linear dissipative system},\ }\href
	{https://doi.org/10.1006/aphy.2000.6017} {\bibfield  {journal} {\bibinfo
			{journal} {Annals of Physics}\ }\textbf {\bibinfo {volume} {281}},\ \bibinfo
		{pages} {547} (\bibinfo {year} {2000})}\BibitemShut {NoStop}%
	\bibitem [{\citenamefont {Ingold}()}]{Ingold2002}%
	\BibitemOpen
	\bibfield  {author} {\bibinfo {author} {\bibfnamefont {G.-L.}\ \bibnamefont
			{Ingold}},\ }\bibfield  {title} {\bibinfo {title} {{Path Integrals and Their
				Application to Dissipative Quantum Systems}},\ }\href@noop {} {\bibfield
		{journal} {\bibinfo  {journal} {Lecture Notes in Physics, Vol. 611, pp. 1-53
				(Springer, New York, 2002)}\ }}\Eprint
	{https://arxiv.org/abs/quant-ph/0208026} {arXiv:quant-ph/0208026 [quant-ph]}
	\BibitemShut {NoStop}%
	\bibitem [{\citenamefont {Giazotto}\ \emph {et~al.}(2006)\citenamefont
		{Giazotto}, \citenamefont {Heikkil{\"a}}, \citenamefont {Luukanen},
		\citenamefont {Savin},\ and\ \citenamefont
		{Pekola}}]{giazotto2006opportunities}%
	\BibitemOpen
	\bibfield  {author} {\bibinfo {author} {\bibfnamefont {F.}~\bibnamefont
			{Giazotto}}, \bibinfo {author} {\bibfnamefont {T.~T.}\ \bibnamefont
			{Heikkil{\"a}}}, \bibinfo {author} {\bibfnamefont {A.}~\bibnamefont
			{Luukanen}}, \bibinfo {author} {\bibfnamefont {A.~M.}\ \bibnamefont
			{Savin}},\ and\ \bibinfo {author} {\bibfnamefont {J.~P.}\ \bibnamefont
			{Pekola}},\ }\bibfield  {title} {\bibinfo {title} {Opportunities for
			mesoscopics in thermometry and refrigeration: Physics and applications},\
	}\href {https://doi.org/10.1103/RevModPhys.78.217} {\bibfield  {journal}
		{\bibinfo  {journal} {Reviews of Modern Physics}\ }\textbf {\bibinfo {volume}
			{78}},\ \bibinfo {pages} {217} (\bibinfo {year} {2006})}\BibitemShut
	{NoStop}%
	\bibitem [{\citenamefont {Saira}\ \emph {et~al.}(2007)\citenamefont {Saira},
		\citenamefont {Meschke}, \citenamefont {Giazotto}, \citenamefont {Savin},
		\citenamefont {M{\"o}tt{\"o}nen},\ and\ \citenamefont
		{Pekola}}]{saira2007heat}%
	\BibitemOpen
	\bibfield  {author} {\bibinfo {author} {\bibfnamefont {O.-P.}\ \bibnamefont
			{Saira}}, \bibinfo {author} {\bibfnamefont {M.}~\bibnamefont {Meschke}},
		\bibinfo {author} {\bibfnamefont {F.}~\bibnamefont {Giazotto}}, \bibinfo
		{author} {\bibfnamefont {A.~M.}\ \bibnamefont {Savin}}, \bibinfo {author}
		{\bibfnamefont {M.}~\bibnamefont {M{\"o}tt{\"o}nen}},\ and\ \bibinfo {author}
		{\bibfnamefont {J.~P.}\ \bibnamefont {Pekola}},\ }\bibfield  {title}
	{\bibinfo {title} {Heat transistor: Demonstration of gate-controlled
			electronic refrigeration},\ }\href
	{https://doi.org/10.1103/PhysRevLett.99.027203} {\bibfield  {journal}
		{\bibinfo  {journal} {Physical review letters}\ }\textbf {\bibinfo {volume}
			{99}},\ \bibinfo {pages} {027203} (\bibinfo {year} {2007})}\BibitemShut
	{NoStop}%
	\bibitem [{\citenamefont {Bell}(2008)}]{bell2008cooling}%
	\BibitemOpen
	\bibfield  {author} {\bibinfo {author} {\bibfnamefont {L.~E.}\ \bibnamefont
			{Bell}},\ }\bibfield  {title} {\bibinfo {title} {Cooling, heating, generating
			power, and recovering waste heat with thermoelectric systems},\ }\href
	{https://doi.org/10.1126/science.1158899} {\bibfield  {journal} {\bibinfo
			{journal} {Science}\ }\textbf {\bibinfo {volume} {321}},\ \bibinfo {pages}
		{1457} (\bibinfo {year} {2008})}\BibitemShut {NoStop}%
	\bibitem [{\citenamefont {Yan}\ and\ \citenamefont {Xu}(2005)}]{Yan2005}%
	\BibitemOpen
	\bibfield  {author} {\bibinfo {author} {\bibfnamefont {Y.}~\bibnamefont
			{Yan}}\ and\ \bibinfo {author} {\bibfnamefont {R.}~\bibnamefont {Xu}},\
	}\bibfield  {title} {\bibinfo {title} {{Quantum Mechanics of Dissipative
				Systems}},\ }\href
	{https://doi.org/10.1146/annurev.physchem.55.091602.094425} {\bibfield
		{journal} {\bibinfo  {journal} {Annual Review of Physical Chemistry}\
		}\textbf {\bibinfo {volume} {56}},\ \bibinfo {pages} {187} (\bibinfo {year}
		{2005})}\BibitemShut {NoStop}%
\end{thebibliography}
\end{comment}

\end{document}